\documentclass[fleqn,usenatbib,useAMS]{mnras}

\usepackage{mathptmx}

\usepackage[T1]{fontenc}

\usepackage{graphicx}	
\usepackage{amsmath}	
\usepackage{amssymb}	

\usepackage{ulem}

\newcommand{\kev}{keV}

\newcommand{\nustar}{\textit{NuSTAR}}
\newcommand{\xmm}{\textit{XMM-Newton}}

\newcommand{\fe}{Fe~K$\alpha$}
\newcommand{\etal}{et al.}
\newcommand{\rexcor}{\textsc{reXcor}}
\newcommand{\he}{HE~1143-1820}
\newcommand{\ngc}{NGC~4593}

\title[\textsc{reXcor}: An X-ray Spectral Model for AGNs]{\textsc{reXcor}: A Model of the X-ray Spectrum of Active Galactic Nuclei that Combines Ionized Reflection and a Warm Corona}

\author[X. Xiang \etal]{
X. Xiang\thanks{E-mail: xxiang37@gatech.edu},$^{1}$ D. R. Ballantyne\thanks{E-mail: david.ballantyne@physics.gatech.edu},$^{1}$ S. Bianchi,$^{2}$ A. De Rosa,$^{3}$ G. Matt,$^{2}$ R. Middei,$^{4,5}$ P.-O. Petrucci,$^{6}$ \newauthor A. R{\'o}{\.z}a{\'n}ska$^{7}$ and F. Ursini$^{2}$
\\
$^1$Center for Relativistic Astrophysics, School of Physics, Georgia
  Institute of Technology, 837 State Street, Atlanta, GA 30332-0430, USA\\
$^2$Dipartimento di Matematica e Fisica, Universit\`{a} degli Studi Roma Tre, via della Vasca Navale 84, 00146 Roma, Italy\\
$^3$INAF – Istituto di Astrofisica e Planetologie Spaziali, Via Fosso del Cavaliere, I-00133 Roma, Italy\\
$^4$INAF – Osservatorio Astronomico di Roma, Via Frascati 33, 00040 Monte Porzio Catone, Italy\\
$^5$Space Science Data Center, SSDC, ASI, Via del Politecnico snc, 00133 Roma, Italy\\
$^6$Univ. Grenoble Alpes, CNRS, IPAG, 38000 Grenoble, France\\
$^7$Nicolaus Copernicus Astronomical Center, Polish Academy of Sciences, Bartycka 18, PL-00-716 Warszawa, Poland
}

\date{Accepted XXX. Received YYY; in original form ZZZ}

\pubyear{2022}

\begin{document}
\label{firstpage}
\pagerange{\pageref{firstpage}--\pageref{lastpage}}
\maketitle

\begin{abstract}
The X-ray spectra of active galactic nuclei (AGNs) often exhibit an excess of emission above the primary power-law at energies $\la 2$~\kev. Two models for the origin of this `soft excess' are ionized relativistic reflection from the inner accretion disc and Comptonization of thermal emission in a warm corona. Here, we introduce \rexcor, a new AGN X-ray ($0.3$--$100$~\kev) spectral fitting model that self-consistently combines the effects of both ionized relativistic reflection and the emission from a warm corona. In this model, the accretion energy liberated in the inner disc is distributed between a warm corona, a lamppost X-ray source, and the accretion disc. The emission and ionized reflection spectrum from the inner $400$~$r_g$ of the disc is computed, incorporating the effects of relativistic light-bending and blurring. The resulting spectra predict a variety of soft excess shapes and sizes that depend on the fraction of energy dissipated in the warm corona and lamppost. We illustrate the use of \rexcor\ by fitting to the joint \xmm\ and \nustar\ observations of the Seyfert 1 galaxies \he\ and \ngc, and find that both objects require a warm corona contribution to the soft excess. Eight \rexcor\ table models, covering different values of accretion rate, lamppost height and black hole spin, are publicly available through the XSPEC website. Systematic use of \rexcor\ will provide insight into the distribution of energy in AGN accretion flows. 
\end{abstract}




\section{Introduction}
\label{sect:intro}
At energies $\la 2$~keV the X-ray spectra of most active galactic nuclei (AGNs) exhibit an excess of emission above what is expected from the primary hard X-ray power-law \citep[e.g.,][]{piconcelli2005,scott12,winter12,ricci17,gw20}. This `soft excess' can be modeled as a thermal emitter \citep[e.g.,][]{tp89,bianchi09}, but the resulting temperatures do not appear to vary with the black hole mass or accretion rates, in contrast with expectations from standard accretion disc theories \citep[e.g.,][]{gd04,bianchi09}. Therefore, the origin of the soft excess is expected to be closely related to local atomic and radiative processes within the accretion disc \citep[e.g.,][]{gd04,crummy06}. The detection of high frequency soft lags from numerous bright Seyferts shows that some fraction of the soft excess originates at a distance of a few gravitational radii ($r_g=GM/c^2$, where $M$ is the black hole mass) from the hard X-ray emitting corona \citep[e.g.,][]{demarco13,kara16,cbk21}.

While these reverberation measurements show that at least some part of the soft excess is produced by the reprocessing of irradiating X-rays (a process known as X-ray reflection; \citealt{fr10}), other processes occurring within the accretion disc are likely to contribute to the soft excess \citep[e.g.,][]{kb16}. Evidence for this additional component can be seen from the difficulties faced by reflection models when fitting broadband X-ray spectra of AGNs \citep[e.g.,][]{matt14,porquet18,laha21,xxu21}. Although the models provide an adequate description of the data, the strength of the soft excess can drive the models to extreme conditions, such as high disc densities, large iron abundances, or small inclination angles \citep[e.g.,][]{crummy06,garcia19,jjiang19,jjiang20,middei20}. Therefore, there has been significant interest in determining other sources for the soft excess that can alleviate the challenges faced by a reflection origin.

In recent years, interest has focused on the idea of a `warm corona' as an alternative origin for the soft excess. A warm corona is a Comptonizing layer at the surface of the accretion disc with a Thomson depth of $\tau \sim 10$--$40$ and temperature $kT \sim 0.1$--$1$~keV \citep[e.g.,][]{mag98,matt14,kd18,petrucci18}. This layer, heated by internal dissipation of accretion energy, would produce the soft excess by scattering the thermal emission from the bulk of the disc as it passes through the warm corona \citep[e.g.,][]{mehdipour15}. Although a straightforward warm Comptonization model provides a good fit to the observed soft excess in many AGNs \citep[e.g.,][]{petrucci18,middei18,middei19,middei20,ursini20}, there are concerns about the physical plausibility of the scenario. For example, \citet{garcia19} argued that thermal disc emission passing through a $\sim 1$~\kev\ gas should be imprinted with many soft X-ray absorption lines, which are not observed in AGN spectra. On the theoretical side, \citet{roz15} and \citet{gr20} showed that significant magnetic pressure support is required to produce a $\tau \sim 10$ warm corona in hydrostatic equilibrium. 

Recently, \citet{ball20} found that the hard X-ray power-law illuminating the surface of a warm corona is crucial to both heating the layer and providing a base level of ionization in the gas. As a result of the X-ray heating and ionization, thermal radiation passing through the warm corona would avoid being lost to soft X-ray absorption lines. \citet{ball20} also showed that a warm corona can produce a smooth soft excess, but only for a limited range of gas densities and temperatures. The implications of these results is that any warm corona must be placed close to the hard X-ray emission region in order to have sufficient ionization, and that changes in the soft excess may closely track structural changes in the warm corona \citep{bx20}. Similar conclusions on the warm corona properties under the conditions of X-ray illumination were found independently by \citet{petrucci20}.

The picture that emerges from these studies is that while relativistic reflection and a warm corona are both plausible origins for the soft excess, the combination of the two scenarios may be a natural output from the inner accretion disc of AGNs \citep[e.g.,][]{porquet18,porquet21,xu21}. In order to explore this idea further, we describe and present in this paper results from \rexcor, a new publicly available, phenomenological AGN X-ray fitting model that self-consistently includes emission from both a warm corona and relativistically blurred ionized reflection. Application of \rexcor\ to X-ray spectral data will show how these two components combine to produce the soft excess, and lead to constraints on how the accretion energy flows between the hot and warm coronae in AGN discs.

We describe the ingredients of \rexcor\ and how it is calculated in the next section. An overview of the resulting spectral model and how the spectra change in response to the model parameters is provided in Sect.~\ref{sect:rexcor}. We illustrate the use of \rexcor\ in Sect.~\ref{sect:fits} where the model is used to fit a series of \xmm\ and \nustar\ spectra from two AGNs, \he\ and \ngc. Finally, a summary of the paper is presented in Sect.~\ref{sect:summary}. Table models of \rexcor\ spectra are available for use by the community at the XSPEC website.

\section{Model Description}
\label{sect:model}
\begin{figure*}
  \includegraphics[width=0.9\textwidth]{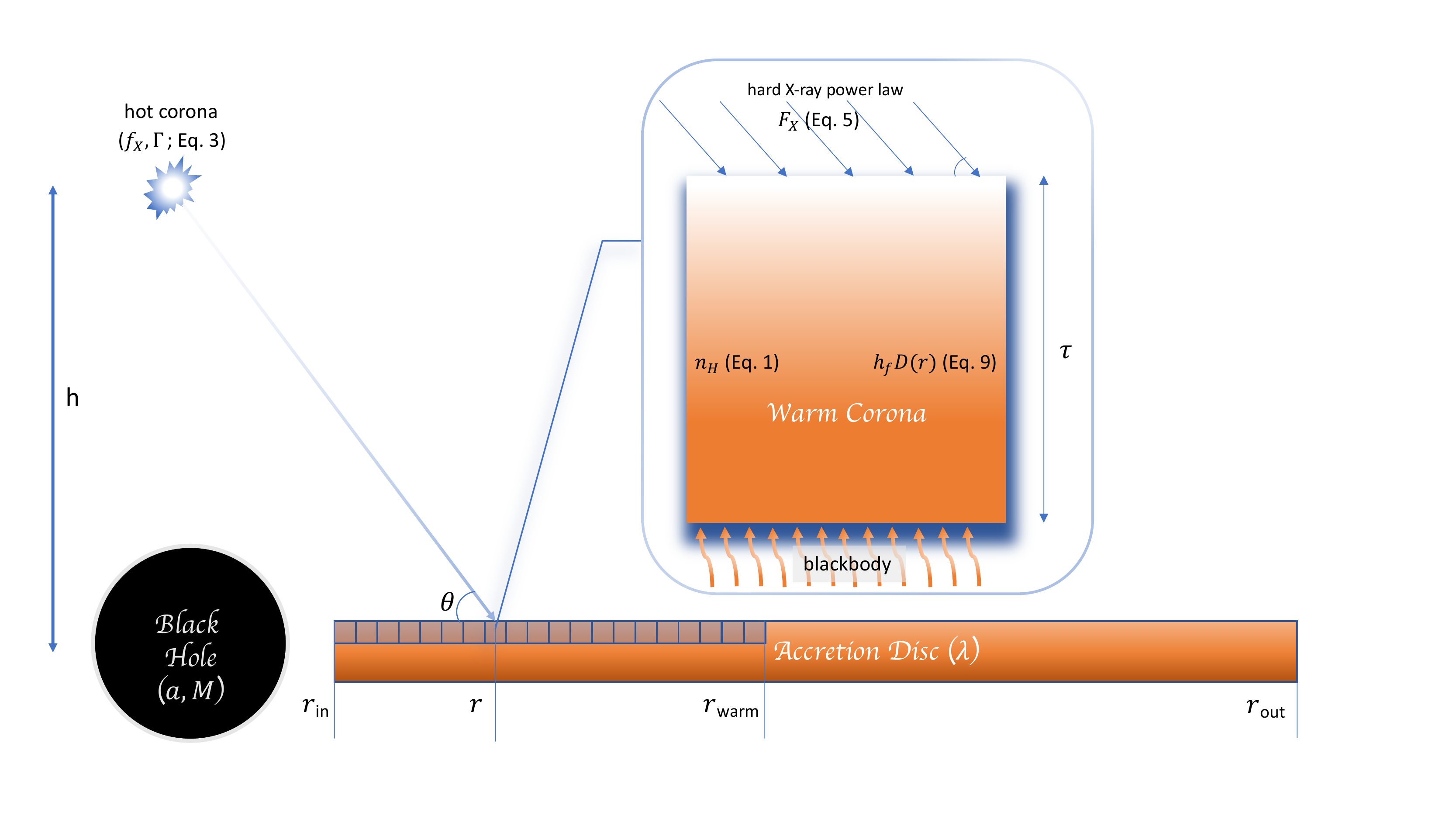}
\caption{A schematic illustration of the ingredients behind the
  calculation of a \rexcor\ spectrum. A hot corona in a lamppost
  geometry is situated at a height $h$ above a black hole of mass $M$
  and spin $a$. The lamppost produces a power-law spectrum (with
  photon-index $\Gamma$) with a luminosity parameterized by $f_X$
  (Eq.~\ref{eq:Lx}). The hard X-ray power-law illuminates the surface
  of a thin accretion disc with accretion rate $\lambda$. The flux at
  the disc surface is given by Eq.~\ref{eq:fx} which incorporates the
  effects of lightbending. We consider a column of irradiated gas with
Thomson depth $\tau$ and constant density $n_{\mathrm{H}}$
(Eq.~\ref{eq:density}). This column is also heated by a warm corona
which dissipates an energy flux $h_f D(r)$ (Eq.~\ref{eq:hf}) in the
gas. In addition to the internal heating and external X-ray
illumination, thermal blackbody radiation from the bulk of the disc
may enter the slab from below. We compute the X-ray reflection and emission
spectrum from this slab at radius $r$, and integrate these spectra
over the disc to produce the final \rexcor\ model, including the
effects of relatvistic blurring (see Sects.~\ref{sub:singlespect}
and~\ref{sub:fulldisk} for complete details).}
\label{fig:schematic}
\end{figure*}

A \rexcor\ model is constructed by integrating the reflection and
emission spectrum produced by an AGN accretion disc from an inner
radius ($r_{\mathrm{in}}=r_{\mathrm{ISCO}}+0.5$\footnote{The addition
of $0.5$ to $r_{\mathrm{ISCO}}$ is to avoid the unphysical steep
increase in disc density predicted by Eq.~\ref{eq:density}.}, where
$r_{\mathrm{ISCO}}$ is the radius of the innermost stable circular
orbit of a prograde accretion disc and all distances are in units of
$r_g$) to an outer radius, $r_{\mathrm{out}}=400$. Below, we first
describe how the spectrum is computed at a specific disc radius $r$,
and then explain the integration procedure to construct the final
spectrum. A schematic diagram of our model set-up, with references to
the equations described below, is shown in Figure~\ref{fig:schematic}.

\subsection{The Reflection and Emission Spectrum at Radius $r$}
\label{sub:singlespect}
The calculation of the X-ray spectrum from a disc radius $r$ that includes the effects of both reflection and a warm corona largely follows the procedure described by \citet{ball20} and \citet{bx20}. We compute the reflection and emission spectrum from a one-dimensional slab located at the surface of an accretion disc with hydrogen number density $n_{\mathrm{H}}$ and Thomson depth $\tau$. To determine the density of the slab, the mid-plane density is computed following the \citet{sz94} radiation-pressure dominated disc solution and then divided by $10^3$ to mimic the fall-off in density from the mid-plane to the surface \citep[e.g.,][]{jiang19}. Assuming $n_{\mathrm{H}}=\rho/m_{\mathrm{p}}$, where $\rho$ is the gas density and $m_{\mathrm{p}}$ is the mass of a proton, the hydrogen number density in the slab is
\begin{equation}
    \label{eq:density}
        n_{\mathrm{H}} = (2.4 \times 10^{15}) \left (\frac{\eta}{0.1} \right )^2 \left (\frac{\alpha}{0.1} \right)^{-1} \left(\frac{M}{\mathrm{M}_\odot}  \right)^{-1} \lambda ^{-2} r^{3/2} J(r)^{-2}\ \mathrm{cm^{-3}},
\end{equation}
where $\eta$ is the radiative efficiency of the accretion process, $\alpha$ is the \cite{ss73} viscosity parameter, and $J(r) = 1-(r_{\mathrm{ISCO}}/r)^{1/2}$. In addition, the Eddington ratio of the AGN is $\lambda=L_{\mathrm{bol}}/L_{\mathrm{Edd}}$, where $L_{\mathrm{bol}}$ is the bolometric luminosity and $L_{\mathrm{Edd}}=4\pi GM m_{\mathrm{p}} c/\sigma_{\mathrm{T}}$ is the Eddington luminosity. Throughout this paper, both $\alpha$ and $\eta$ are fixed at $0.1$. 
For simplicity, we remove the dependence on the black hole spin in Eq.~\ref{eq:density} by assigning $r_{\mathrm{ISCO}}=1.45$ when calculating $n_{\mathrm{H}}$. As a result, for a given black hole mass the disc density depends only on $\lambda$ and $r$.

The slab of gas at $r$ is illuminated from above by a stationary X-ray emitting hot corona that is located at a height $h$ (in units of $r_g$) above the rotational axis of the black hole (i.e., a lamppost geometry; e.g., \citealt{matt91,mm96,dauser13}). This geometry is consistent with recent reverberation results indicating that the corona must be compact and located close to the central black hole \citep{demarco13,kara16}. The hot corona emits a power-law spectrum with photon index $\Gamma$ (i.e., the photon flux $\propto E^{-\Gamma}$) and exponential cutoff energies at both $30$~eV and $300$~keV to account for its Comptonization origin \citep[e.g.,][]{petrucci01}. The highest energy considered by the \citet{brf01} code is 98~\kev, so the precise value of the high-energy roll-over has minimal effect on the results\footnote{Including the effects of photons at energies $> 98$~keV would lead to higher gas temperatures at the surface ($\tau \la 1$) of the slab, in particular for hard ($\Gamma < 2$) X-ray power-laws \citep{garcia13}. As this effect is concentrated at the surface of the slab, it does not replace the heating provided by a warm corona.}.

Each side of the disc produces an energy flux $D(r)$ due to dissipation within the disc \citep{ss73} where
\begin{equation}
    D(r) = (6.89\times 10^{27}) \left (\frac{\eta}{0.1} \right )^{-1} \left(\frac{M}{\mathrm{M}_\odot}\right)^{-1} \lambda r^{-3}J(r)\ \mathrm{erg\ cm^{-2}\ s^{-1}.}
    \label{eq:D}
\end{equation}
The total X-ray luminosity of the hot corona is related to a constant fraction $f_X$ of $D(r)$ within a critical radius $r_{\mathrm{c}}$:
\begin{equation}
\label{eq:Lx}
    L_X = (1.50\times 10^{38}) \left (\frac{\eta}{0.1} \right )^{-1} \lambda \left(\frac{M}{\mathrm{M}_\odot}\right)\int_{r_{\mathrm{ISCO}} }^{r_{\mathrm{c}}}f_X r^{-3} J(r) dS(r)\ \mathrm{erg\ s^{-1}},
\end{equation}
where
\begin{equation} \label{eq:dS}
dS  = 2 \pi r \sqrt{\frac{r^2+a^2+2a^2/r}{r^2-2r+a^2}} dr
\end{equation}
is the proper element of the disc area at the midplane \citep{vincent2016} and $a$ is the black hole spin. The critical radius is fixed at $r_c=10$ for all models, consistent with the observational evidence for a highly compact corona \citep[e.g.,][]{rm13}, meaning that regions of the disc at $r > r_c$ do not contribute to the hot corona. As $D(r)$ falls rapidly with $r$, increasing $r_c$ beyond $10$ has a negligible effect on the final model spectra.

The X-ray flux at the surface of the disc at $r$ is affected by light-bending, and therefore depends on $h$ and $a$ \citep[e.g.,][]{mf04,fk07,dauser13}. The flux at radius $r$ is \citep{ball17}
\begin{equation}
    \label{eq:fx}
    F_X(r) = \frac{L_XF(r,h)g_{\mathrm{lp}}^2}{z(M)A},
\end{equation}
where $F(r,h)$ is given by the \citet{fk07} fitting formulas for the illumination pattern on the accretion disc,
\begin{equation}
    \label{eq:glp}
    g_{\mathrm{lp}}=
{r^{3/2} + a \over \sqrt{r^3+2ar^{3/2}-3r^2}} \sqrt{{h^2 + a^2 -2h}
  \over h^2 +a^2}
\end{equation}
is the ratio of the photon frequency at the disc to the frequency at the X-ray source \citep{dauser13},
\begin{equation}
 \label{eq:zofM}
z(M)=\left ( {GM_{\odot} \over c^2} \right )^2 \left ( {M \over
    M_{\odot}} \right )^2
\end{equation}
converts the area to physical units \citep{ball17}, and
\begin{equation}
    \label{eq:A}
    A = \int_{r_{\mathrm{ISCO}}}^{r_{\mathrm{out}}}F(r,h) g_{lp}^2 dS(r)
\end{equation}
is a normalization factor to ensure the total flux on the disc integrates to $L_X$. 
The fluxes computed this way are valid for $r > 1.15$ and $3 \leq h \leq 100$ because of the use of the \citet{fk07} fitting functions.

Light-bending also strongly influences the incident angle, $\theta_i(r)$, of the radiation on the disc \citep[e.g.,][their Fig. 5]{dauser13} where radii closer to the black hole are generally irradiated at smaller angles, but most of the disc is illuminated at large $\theta_i$. We approximate this effect using a straightforward Newtonian description: $\tan \theta_i = r/h$. 

%

%


To include the effects of heating from a warm corona in our calculation, an energy flux $h_f D(r)$ is assumed to be uniformly distributed throughout the constant density slab, which corresponds to a heating function (in erg cm$^3$ s$^{-1}$) of 
\begin{equation}
  \label{eq:hf}
    \mathcal{H} = \frac{h_fD(r)\sigma_{\mathrm{T}}}{\tau n_{\mathrm{H}}}.
\end{equation}
This function is equivalent to a constant heating rate per particle over $\tau$, and $n_{\mathrm{H}} n_e \mathcal{H}$ is the heating rate per unit volume added to the thermal balance equation \citep[e.g.,][their Eq. 7]{ross79}. 

Finally, the remaining energy flux at $r$, $(1-f_X-h_f)D(r)$ when $r < r_c$, is injected as a blackbody into the lowest zone of the slab with a temperature given by the standard blackbody relationship. For $r > r_c$, the energy flux released in the lower zone is $(1-h_f)D(r)$.

At this point, we can compute the rest-frame reflection and emission spectrum from the top $\tau$ of an accretion disc at radius $r$ due to irradiation from the lamppost from above, the blackbody from below, and warm corona heating distributed throughout the layer \citep[e.g.,][]{ball20,bx20}. The calculation solves the thermal and ionization balance of the layer, and includes cooling lines from C~\textsc{v-vi}, N~\textsc{vi-vii}, O~\textsc{v-viii}, Mg~\textsc{ix-xii}, Si~\textsc{xi-xiv} and Fe~\textsc{xvi-xxvi} assuming Solar abundances \citep[e.g.][]{rf93,rfy99,brf01}. Comptonization of the outward diffuse X-rays, which includes the blackbody emission and scattered photons from the hard power-law, is transferred using a Fokker-Planck operator \citep{rwm78,ross79} and is sensitive to the thermal structure of the gas determined by heating from both the hot and warm coronas. Therefore, any soft excess in the resulting reflection and emission spectrum self-consistently includes the effects of both reflection and the warm corona.

\subsection{Constructing the Integrated Spectrum}
\label{sub:fulldisk}
The final model spectrum is calculated by integrating the spectra computed at different $r$ from $r_{\mathrm{in}}$ to $r_{\mathrm{out}}$. To do the integration efficiently, we separate the disc into two regions based on the value of the ionization parameter of the disc,
\begin{equation} \label{eq:xi}
    \xi (r) = \frac{4\pi F_{X}(r)}{n_{\mathrm{H}}(r)}.
\end{equation}
Inside a radius $r_{\mathrm{warm}}$, defined by $\xi(r_{\mathrm{warm}})=5$~erg~s~cm$^{-1}$, the disc is strongly irradiated and will have a significant surface ionization gradient. We split the range $r_{\mathrm{in}} \leq r \leq r_{\mathrm{warm}}$ into 20 annuli with a step size $dr_i=(r_{\mathrm{warm}}-r_{\mathrm{in}})/20$, and use the method described above to compute the spectrum from each annulus. 
However, the disc is weakly illuminated beyond $r_{\mathrm{warm}}$ and the reflection spectrum is dominated by low-ionization lines and edges, such as a strong \fe\ line at 6.4~keV. Since the the shape of the reflection spectrum will be dominated by these low ionization features at $r > r_{\mathrm{warm}}$, the spectrum calculated at $r_{\mathrm{warm}}$ (scaled to match the appropriate $F_X(r)$) is used for all radii in the range $r_{\mathrm{warm}} < r \leq r_{\mathrm{out}}$ with a step size $dr_o = 5$. 

\begin{figure*}
\includegraphics[width=0.9\textwidth]{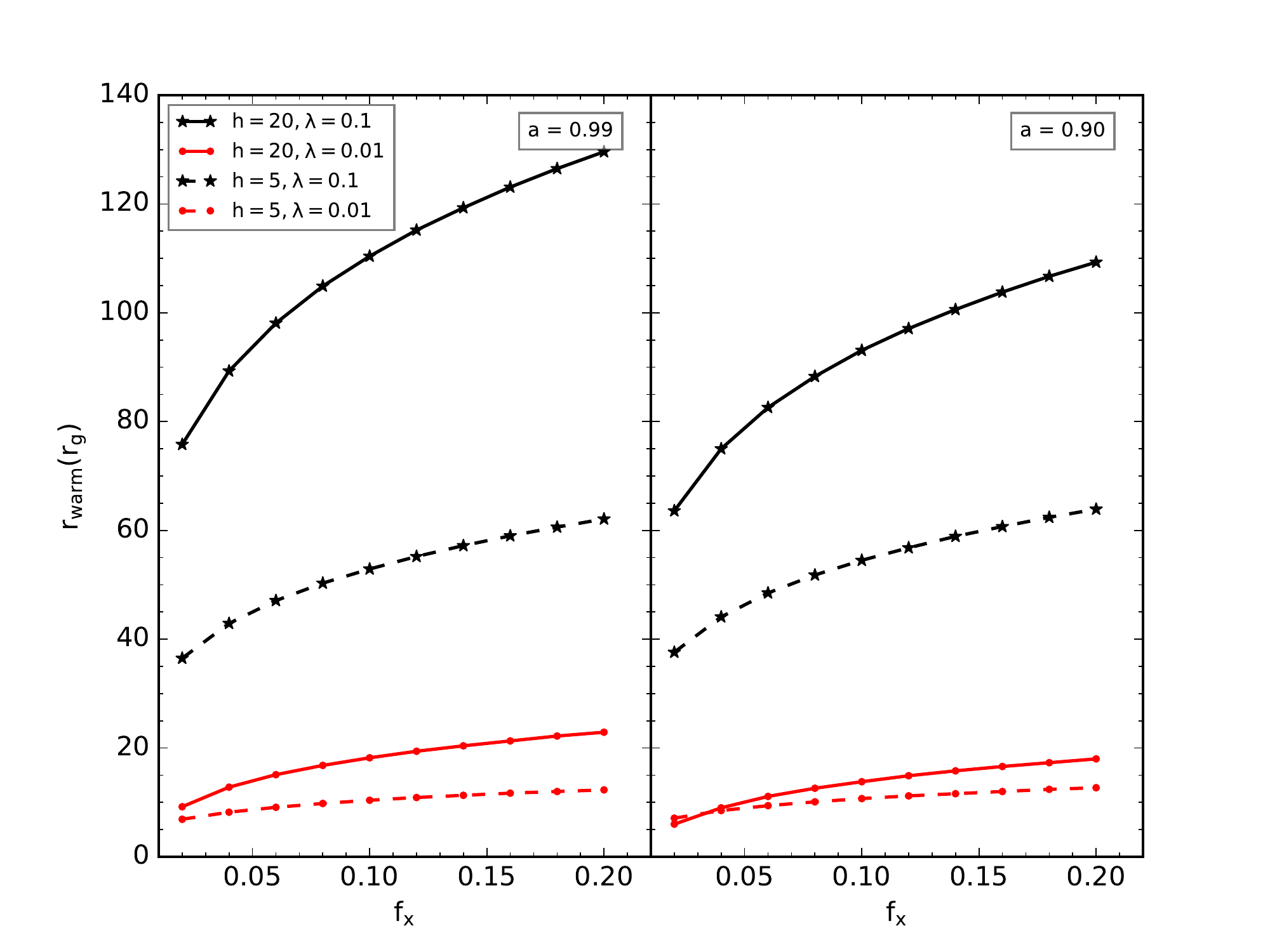}
\caption{The value of $r_{\mathrm{warm}}$ as a function of the X-ray heating fraction $f_X$ for different values of $h$, $\lambda$ and $a$. The reflection and emission spectrum evaluated at $r_{\mathrm{warm}}$ is used to extend the radial integration to $r_{\mathrm{out}}$. For $r < r_{\mathrm{warm}}$ the emission and reflection spectrum is computed for 20 annuli between $r_{\mathrm{in}}$ and $r_{\mathrm{warm}}$.  As $r_{\mathrm{warm}}$ is determined by the condition $\xi(r_{\mathrm{warm}})=5$~erg~s~cm$^{-1}$, it is increased by larger values of $f_X$ and $\lambda$ (which increases the X-ray flux on the disc), and decreased by a lower $h$ (where light-bending focuses the flux onto the inner disc). There is also a modest drop in $r_{\mathrm{warm}}$ with $a$, in particular at $h=20$, due to a slight decrease in the X-ray luminosity (Eq.~\ref{eq:Lx}). There is a weak dependence of $r_{\mathrm{warm}}$ on $h_f$ and $\tau$ when $\Gamma > 2$ and $\lambda=0.1$, which is not shown in the Figure. In this case, $r_{\mathrm{warm}}$ is lowered by an average of $3.75$\% from what is plotted here with a maximum decrease of $12$\%.}
\label{fig:rwarm}
\end{figure*}

Figure~\ref{fig:rwarm} shows how the values of $r_{\mathrm{warm}}$ are impacted by different parameters of the model.
In all cases, $r_{\mathrm{warm}}$ increases with $f_X$, as this parameter is proportional to the flux irradiating the disc (Eq.~\ref{eq:Lx}). However, the slope of this relation is smaller for lower values of $\lambda$ (comparing the stars and circles) and $h$ (comparing the solid and dotted lines). In the former case, the lower $\lambda$ significantly enhances the disc density (Eq.~\ref{eq:density}) which is not compensated by the changes in $f_X$. Therefore, $r_{\mathrm{warm}}$ is both smaller and less sensitive to $f_X$ for low $\lambda$. In contrast, changes in the lamppost height $h$ affect the radiation pattern on the accretion disc. When $h=5$, for example, light-bending focuses a large fraction of the flux onto the inner accretion disc, and the flux at larger radii is reduced \citep[e.g.,][]{dauser13} leading to a smaller $r_{\mathrm{warm}}$ even for large values of $f_X$. A larger $h$ provides a more uniform illumination of the disc and so $r_{\mathrm{warm}}$ is largest in this situation. The two panels in Fig.~\ref{fig:rwarm} also show that $r_{\mathrm{warm}}$ has a moderate dependence on the black hole spin $a$, in particular when $h=20$ (solid lines), due to $L_X$ having a weak dependence on $a$ (Eq.~\ref{eq:Lx}).

These values of $r_{\mathrm{warm}}$ broadly separate the disc into regions of ionized refection (when $r < r_{\mathrm{warm}}$) and neutral reflection (when $r > r_{\mathrm{warm}}$). Thus, we expect models with low $r_{\mathrm{warm}}$ (i.e., low $\lambda$) to produce spectra that are dominated by neutral reflection features while those with large $r_{\mathrm{warm}}$ will be dominated by highly ionized reflection. Many models, however, will have a mixture of both due to the ionization gradient across the disc. Since $r_{\mathrm{warm}}$ is determined by the X-ray flux and the disc density, it is nearly independent of the warm corona parameters $\tau$ and $h_f$. There is a slight dependence of $r_{\mathrm{warm}}$ on $h_f$ and $\tau$ in the $\lambda=0.1$ models that occurs when $\Gamma > 2$ and the ionizing power of the irradiating spectrum is weakened. In this regime, $r_{\mathrm{warm}}$ can be reduced from the plotted values by an average of $3.75$\%, with a maximum change of $12$\%.

Prior to performing the radial integration, each individual spectrum
from a disc annulus (from $r$ to $r+dr_i$ if $r \leq
r_{\mathrm{warm}}$, or from $r$ to $r+dr_o$ if $r >
r_{\mathrm{warm}}$) is blurred using the \textsc{relconv\_lp}
convolution model \citep{dauser13} to take into account the
relativistic effects seen by a distant observer\footnote{As our
innermost radial zone is at $r_{\mathrm{ISCO}}+0.5$, the amount of
blurring in the final spectrum will be moderately
underestimated for a given value of $a$. Any spin estimates obtained by
\rexcor\ should be considered approximate lower-limits (see also Sect.~\ref{sect:rexcor}).}. The \textsc{relconv\_lp} model is passed the same values of $h$, $a$ and $\Gamma$ as the reflection calculation described in Sect.~\ref{sub:singlespect}. We assume isotropic limb darkening and a disc inclination to the line of sight of $i = 30^{\circ}$. The blurred spectra are then each multiplied by the proper area of the appropriate annulus using Eq.~\ref{eq:dS} and then summed to produce the final \rexcor\ model in units of ergs~s$^{-1}$. 

Before examining the resulting spectra in detail, it is important to recognize the limitations of the method presented here. Although the calculation of the spectrum from each annulus extends to $\approx 1$~eV in order to conserve energy \citep[e.g.,][]{ball20}, the limited number of elements and low-ionization states treated in the code restricts the accuracy of the predicted spectra at energies $\la 0.1$~keV. In addition, the integrated \rexcor\ model uses the spectrum at $r_{\mathrm{warm}}$ for $r > r_{\mathrm{warm}}$. Therefore, the thermal emission from the disc at these radii will not be properly included in the final spectrum. As a result, the \rexcor\ model should only be used in the X-ray band, at energies $\ga 0.3$~keV. We focus on this energy range in the remainder of the paper. Lastly, the lamppost geometry assumed here is only one possible geometry for the location of the X-ray emitting corona. Other geometries, in particular ones with a truncated accretion disk \citep[e.g.][]{petrucci13,kd18}, will predict a different ratio of reflection and warm corona emission in the accretion disc spectra than what is produced by \rexcor. 

\subsection{The reXcor grids}
\label{sub:grids}
The procedure described above produces a single \rexcor\ spectrum given eight parameters: the black hole mass $M$, spin $a$ and accretion rate $\lambda$; the lamppost height $h$ and heating fraction $f_X$; the photon index $\Gamma$, and the warm corona heating fraction $h_f$ and optical depth $\tau$. As our goal is to construct grids of models to fit to broadband X-ray data, this number of parameters is too large to be practical. In addition, it is not physically plausible for parameters such as $M$ and $\lambda$ to be realistically measured with such a phenomenological model using only X-ray data. Therefore, it is worthwhile examining the dependence of the final spectra on the model parameters to determine which are less important and can be removed from a fitting procedure.

\begin{figure*}
\includegraphics[width=0.9\textwidth]{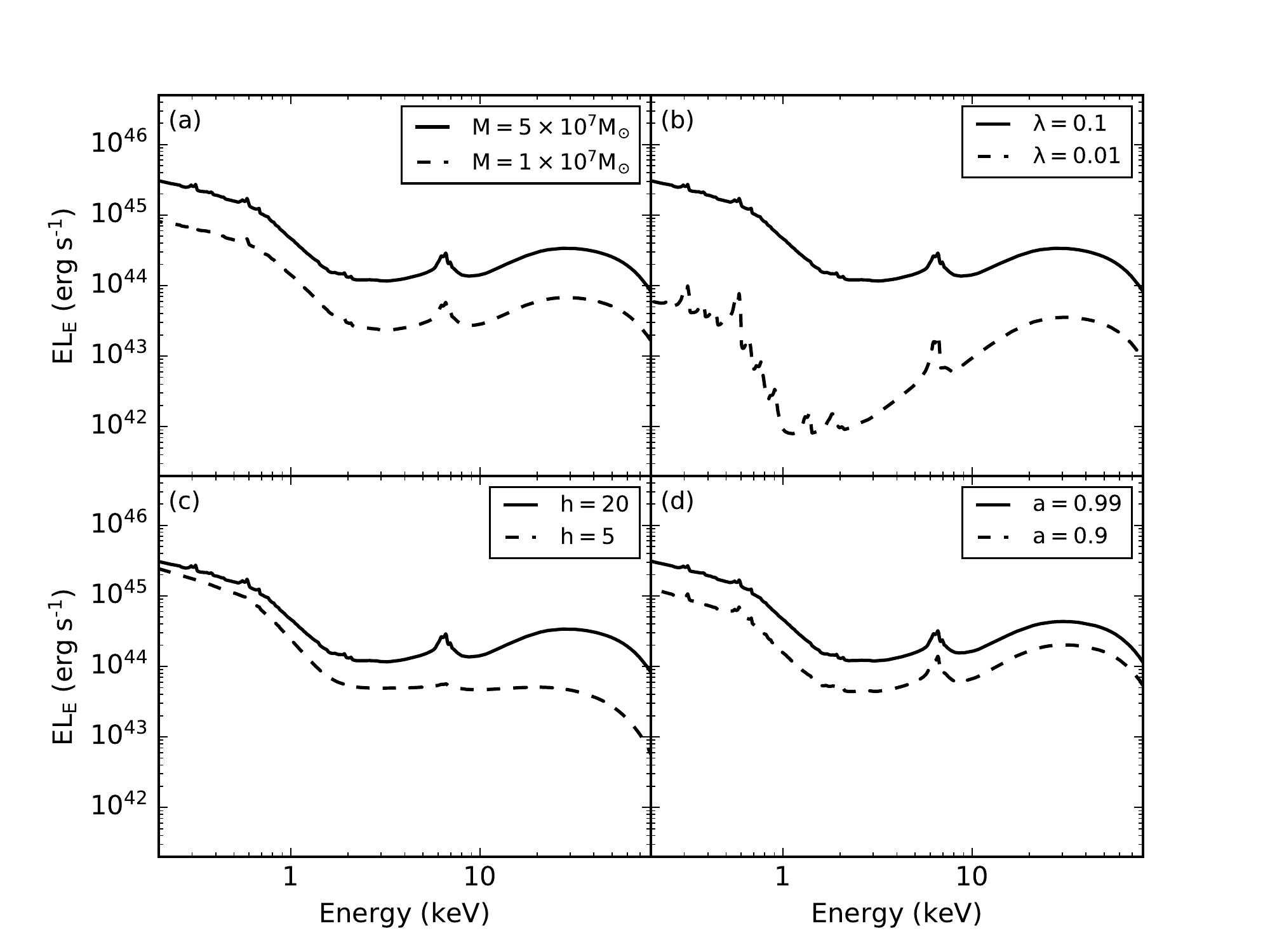}
\caption{Examples of how a \rexcor\ spectrum changes with black hole mass (a), Eddington ratio (b), lamppost height (c) and black hole spin (d). In each panel the solid line plots a \rexcor\ model with the following parameters: $M=5\times 10^7$~M$_{\odot}$, $\lambda=0.1$, $h=20$, $a=0.99$, $h_f=0.4$, $f_X=0.1$, $\tau=20$ and $\Gamma=1.9$. Panel (a) shows that the assumed black hole mass only alters the normalization of the \rexcor\ spectrum. Therefore, the black hole mass is fixed at $5\times 10^7$~M$_{\odot}$ in all \rexcor\ models. Both $\lambda$ and $h$ significantly affect the shape of the model, but can be broadly separated into "high" and "low" cases (i.e., $\lambda=0.1$, $h=20$ and $\lambda=0.01$, $h=5$). The black hole spin produces a small impact on the amplitude of the \rexcor\ spectrum, and leads to minor changes in the spectral shape due to changes in the relativistic blurring. As a result, \rexcor\ grids are only calculated for $a=0.99$ and $a=0.9$.}
\label{fig:bhmass}
\end{figure*}

Figure~\ref{fig:bhmass} shows how a \rexcor\ model depends on four parameters: $M$, $\lambda$, $h$ and $a$. The solid line in each panel shows the same model with $M=5\times 10^7$~M$_{\odot}$, $\lambda=0.1$, $h=20$, $a=0.99$, $h_f=0.4$, $f_X=0.1$, $\Gamma=1.9$ and $\tau=20$.
We see that changes in the black hole mass $M$ (panel a) impact the normalization of the \rexcor\ spectrum, but has a negligible effect on its shape. This is because the spectral shape is largely determined by the ionization parameter $\xi$ (Eq.~\ref{eq:xi}) which, in this model, is independent of $M$ \citep{ball17}. The change in normalization arises from converting distances from units of $r_g$ to physical units, so a smaller $M$ has a physically smaller disc and produces a less luminous spectrum. Since the black hole mass will not impact the determination of the warm corona and reflection parameters (e.g., $h_f$ or $f_X$) $M$ is fixed at $5\times 10^{7}$~M$_{\odot}$ for all \rexcor\ models.

In contrast to the black hole mass, Fig.~\ref{fig:bhmass}(b,c) show that both $\lambda$ and $h$ significantly affect the final \rexcor\ spectrum. A lower $\lambda$ not only means a smaller X-ray luminosity (Eq.~\ref{eq:Lx}), but it also leads to a denser disc (Eq.~\ref{eq:density}) and therefore a much smaller $\xi$ (Eq.~\ref{eq:xi}). Similarly, decreasing the lamppost height from $h=20$ to $h=5$ greatly increases the illumination of the inner accretion disc due to the effects of light-bending \citep[e.g.,][]{fk07,dauser13}. Therefore, the disc is much more ionized in the $h=5$ case which leads to very weak reflection features \citep{ball17}. However, rather than having $\lambda$ and $h$ as two additional fit parameters, we produce models that consider "high" and "low" values in both cases. Therefore, \rexcor\ grids are calculated for either $h=20$ or $h=5$. By performing fits with both sets of models, one may be able to determine if the data are best described by a large or a small lamppost height. Likewise, we produce \rexcor\ models for $\lambda=0.1$ or $\lambda=0.01$, as one of these two values should be applicable for a wide range of AGNs \citep[e.g.,][]{vf07,duras20}.

Lastly, panel (d) of Fig.~\ref{fig:bhmass} shows that the black hole
spin has a relatively minor impact on the \rexcor\ model. The major
effect of spin would be on the ISCO radius of the accretion disc and
the level of relativistic blurring suffered by the emitted
spectrum. However, when the inner disc is ionized (as is the case in
Fig.~\ref{fig:bhmass}), then the changes in relativistic blurring on
the spectrum are very minor. There is also a small drop in
normalization when $a$ is lower due to the dependence of $a$ on the
proper area element (Eq.~\ref{eq:dS}). Again, to keep the size of the
grids manageable, this initial release of \rexcor\ contains grids for
$a=0.99$ and $a=0.9$ as these values bracket the majority of spins
determined in bright AGNs \citep{rey21}.

Table~\ref{table:grids} provides a summary of the 8 \rexcor\ grids
that are publicly available for use in AGN spectral fitting. The grids contain only the reflection and emission spectra from the
accretion disc model described above (see Figs.~\ref{fig:l01h20a099}
and~\ref{fig:l001h20a099} below). A separate
power-law component (with photon-index tied to the \rexcor\ value)
must be included when fitting these models to AGN X-ray data to
account for the illuminating spectrum (Sect.~\ref{sect:fits}). 
\begin{table*}
    \centering
    \caption{There are eight available \rexcor\ grids calculated for two different values of Eddington ratio, black hole spin and lamppost height. The parameters of all grids are listed in Table~\ref{table:params}. A black hole mass $M=5\times 10^7$~M$_{\odot}$ is used in all grids. These table models can be downloaded from the XSPEC website.}  
    \label{table:grids}
    \begin{tabular}{l|c|c|c}
    \hline
        Eddington ratio ($\lambda$) & Black hole spin ($a$) & Lamppost height ($h$) & Filename\\
        \hline
         $0.01$ & $0.9$ & $5$~$r_g$ & \texttt{reXcor\_l001\_a09\_h5.fits} \\
          & & $20$~$r_g$ & \texttt{reXcor\_l001\_a09\_h20.fits} \\
          & $0.99$ & $5$~$r_g$ & \texttt{reXcor\_l001\_a099\_h5.fits} \\
          & & $20$~$r_g$ & \texttt{reXcor\_l001\_a099\_h20.fits} \\
         $0.1$ & $0.9$ & $5$~$r_g$ & \texttt{reXcor\_l01\_a09\_h5.fits} \\
          & & $20$~$r_g$ & \texttt{reXcor\_l01\_a09\_h20.fits} \\
          & $0.99$ & $5$~$r_g$ & \texttt{reXcor\_l01\_a099\_h5.fits} \\
          & & $20$~$r_g$ & \texttt{reXcor\_l01\_a099\_h20.fits} \\
          \hline
          \end{tabular}
\end{table*}

\section{The rexcor spectral Model}
\label{sect:rexcor}
Each of the eight \rexcor\ grids contain 20570 individual spectral models spanning a broad range of $f_X$, $\Gamma$, $h_f$ and $\tau$ (Table~\ref{table:params}). This section describes how these different parameters, which characterize the warm corona and X-ray emitting lamppost in the model, impact the properties of the soft excess and other features in the \rexcor\ spectra. We focus here on the results using the $a=0.99$ grids, and the equivalent figures for two of the $a=0.90$ grids are presented in Appendix~A.
\begin{table} 
	\centering
	\caption{The \rexcor\ grid parameters. The values of $f_X$ covers the range of bolometric corrections observed in many AGNs \citep{vf07,vf09,duras20}. The range of $h_f$ allows for models with no warm corona heating ($h_f=0$) or with dissipationless discs ($h_f=0.8$ and $f_X=0.2$). }
    \label{table:params}
	\begin{tabular}{|l|c|c|c|}
	    \hline
		Parameter & Description & Range & Step Size\\
       \hline
       $f_X$ & lamppost heating fraction & $0.02\ldots0.2$ & $0.02$\\
  $\Gamma$ & photon index of irradiating power-law & $1.7\ldots2.2$ & $0.05$\\
        $h_f$ & warm corona heating fraction & $0.0\ldots0.8$ & $0.05$ \\
        $\tau_T$ & warm corona Thomson depth &  $10\ldots30$ & $2.0$\\
	    \hline
	\end{tabular}
\end{table}

\begin{figure*}
\includegraphics[width=0.825\textwidth]{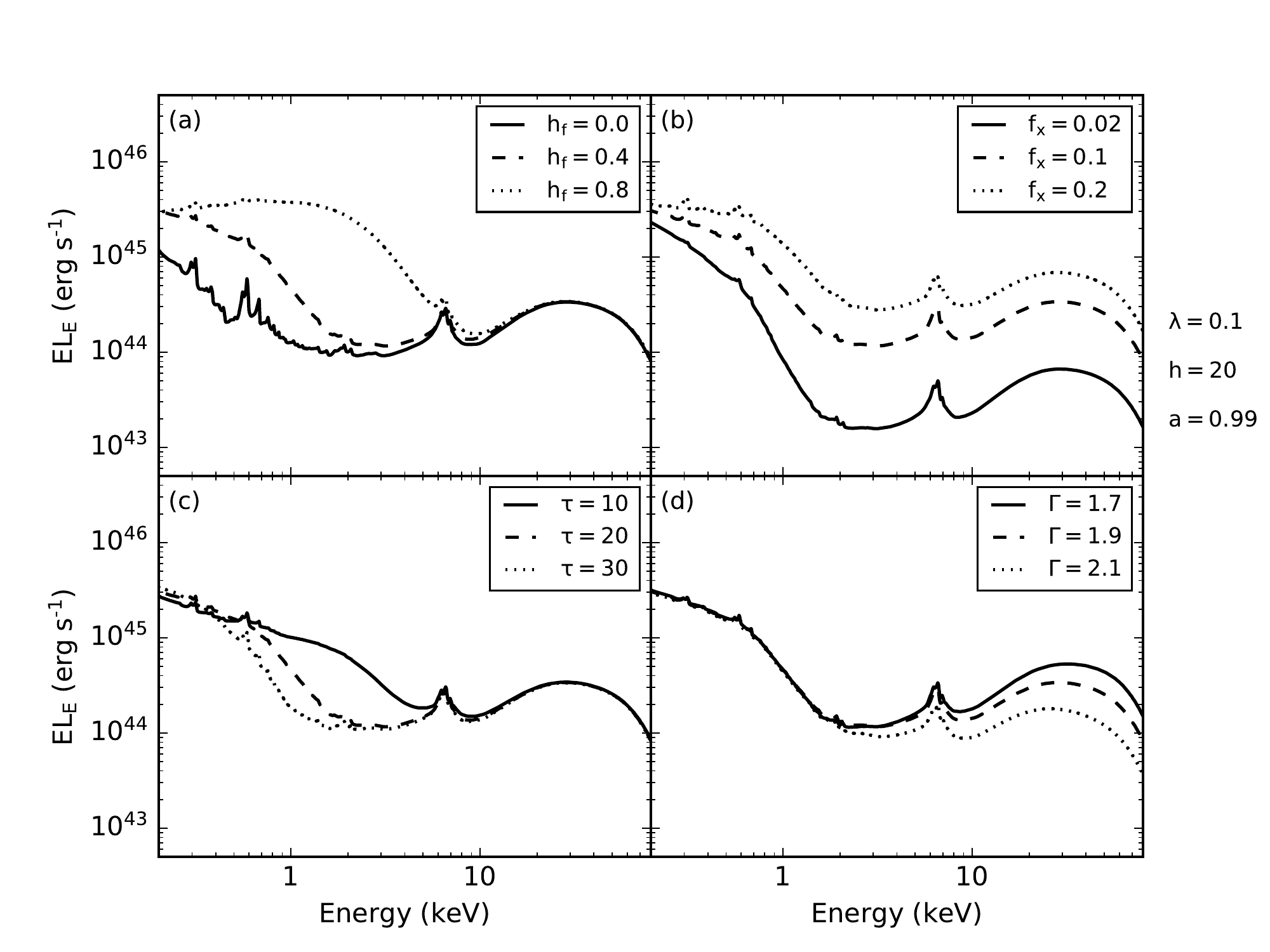}
\includegraphics[width=0.825\textwidth]{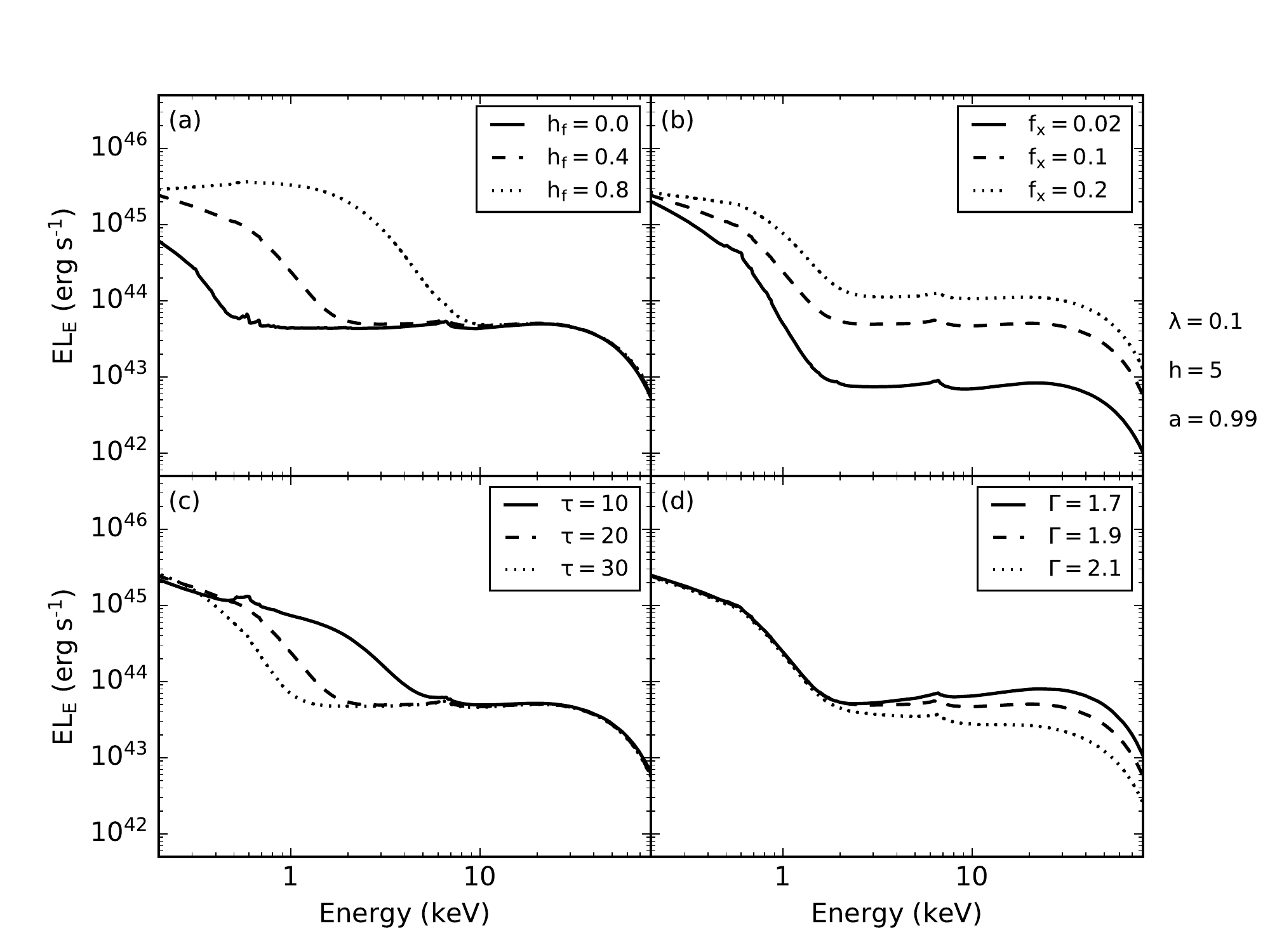}
\caption{(Top) Examples of \rexcor\ spectra from the grid with $\lambda=0.1$, $h=20$ and $a=0.99$. The dashed line in each panel plots the same spectrum with $h_f=0.4$, $f_X=0.1$, $\tau=20$, and $\Gamma=1.9$. The different panels show how the spectrum changes to each of these four parameters. A stronger soft excess can be obtained by increasing $h_f$ or by decreasing $\tau$ or $f_X$. For this value of $\lambda$ and $h$, significant ionized reflection is expected for most spectra. (Bottom) As above, but the spectra are taken from the grid with $\lambda=0.1$ $h=5$ and $a=0.99$. The lower lamppost height produces significant light-bending that enhances the illuminating flux onto the inner accretion disc. The resulting \rexcor\ spectra are dominated by emission from the highly ionized inner accretion disc.}
\label{fig:l01h20a099}
\end{figure*}

The top half of Figure~\ref{fig:l01h20a099} plots several example \rexcor\ spectra from the $\lambda=0.1$, $h=20$, $a=0.99$ grid. 
In each panel the dashed line plots the same spectrum with $h_f=0.4$, $f_X=0.1$, $\tau=20$, and $\Gamma=1.9$, while the other lines show how this baseline model changes due to variations in these four parameters. As expected from Fig.~\ref{fig:rwarm}, all the spectra in this grid show features associated with ionized reflection, such as an ionized \fe\ line and weak emission features at energies $\la 1$~keV. However, variations in the model parameters can lead to major changes to the overall spectral shape. Panel (a) shows that the strength and smoothness of the soft excess is significantly tied to the value of the warm corona heating fraction $h_f$. When $h_f=0$, there is no warm coronal heating in the irradiated disc surface, so all the accretion flux is distributed between the thermal blackbody and the X-ray emitting lamppost. In this case, the soft excess is entirely produced by reflection and exhibits emission and absorption features. However, as $h_f$ is increased, the gas throughout the layer is heated and maintains a higher ionization state \citep{ball20}. This hotter gas enhances the bremsstrahlung emission from the disc, as well as Comptonizing the blackbody radiation emanating from below, leading to a stronger and smoother soft excess. When $h_f=0.8$, the maximum considered in the model, the heating in the disc surface is so large that a Comptonized bremsstrahlung spectrum develops. Thus, the value of $h_f$ can lead to a wide variety of soft excess strengths.

The strength of the reflection signal in a \rexcor\ spectrum is driven
by $f_X$, the hard X-ray heating fraction. As seen in
Fig.~\ref{fig:l01h20a099}(b), changing $f_X$ by an order of magnitude
while $h_f$ is fixed at $0.4$ has the largest impact at energies $\ga
1$~keV. This means that the relative strength of the soft excess can
be reduced by an increase in $f_X$. At low values of $f_X$, the
irradiated gas is less ionized, leading to significant absorption
above 1~keV. This absorption is reduced as the gas becomes further
ionized at larger values of $f_X$ \citep{rfy99}, leading to a weaker
contrast between the soft excess and the higher energy emission. The
constant $h_f=0.4$ and $\tau=20$ ensures that Comptonization is
important in smoothing out features in the soft excess. This panel
also shows that changes in $f_X$ can be approximated as simply varying
the amplitude of the \rexcor\ model. Therefore, because the black hole
spin also changes the normalization of the \rexcor\ spectra
(Fig.~\ref{fig:bhmass}(d)), there is a moderate degeneracy between
changes in $f_X$ and the black hole spin $a$, in the sense that both
parameters can adjust the amplitude of the spectrum. For some datasets
lacking an independent spin estimate (in particular, those with
ionized spectra with few spectral features),
statistically similar fits can be obtained with models that have
lower spin, but higher $f_X$ and models with higher spin, but lower
$f_X$. With high quality data, this degeneracy could be broken, but, in
general, we advise against using \rexcor\ to measure black hole spin.

The optical depth of the warm corona heating layer, $\tau$, also impacts the soft excess (Fig.~\ref{fig:l01h20a099}(c)). As the heat injected into the warm corona is fixed, a smaller $\tau$ will spread the heat into a thinner layer, leading to a stronger increase in temperature. A larger $\tau$, in contrast, will yield a cooler corona since the heating is dissipated over a thicker column of gas. These effects can be seen in the three spectra plotted in this panel, as the $\tau=10$ spectrum shows the effects of Comptonization from a hotter gas.

Panel (d) of Fig.~\ref{fig:l01h20a099} illustrates that the photon index $\Gamma$ does not affect the soft excess in the \rexcor\ models, but is crucial in describing the overall hard X-ray spectral shape. Taken together, the combination of these four parameters allows for a wide range of potential AGN spectral shapes, and, through $h_f$, quantify the contribution of any warm corona in the X-ray spectrum.

The same patterns in the spectra can be seen in the other \rexcor\ grids, but the spectra are qualitatively different due to changes in the illumination conditions. For example, the bottom half of Fig.~\ref{fig:l01h20a099} plots the same sequence of models as the upper half, but now the lamppost height has been reduced to $h=5$. In this scenario, the extreme light-bending that results from the low lamppost height focuses a large fraction of the flux onto the inner accretion disc making it extremely ionized. The radiation pattern greatly enhances the flux from the inner disc and so the integrated spectrum is dominated by the highly ionized inner regions of the disc. As a result, all the spectra produce a weak ionized \fe\ line that is broadened by both relativistic blurring and strong Comptonization. The $h_f=0$ spectrum in panel (a) shows that even with no warm corona, this model produces a smooth soft excess. 

\begin{figure*}
\includegraphics[width=0.815\textwidth]{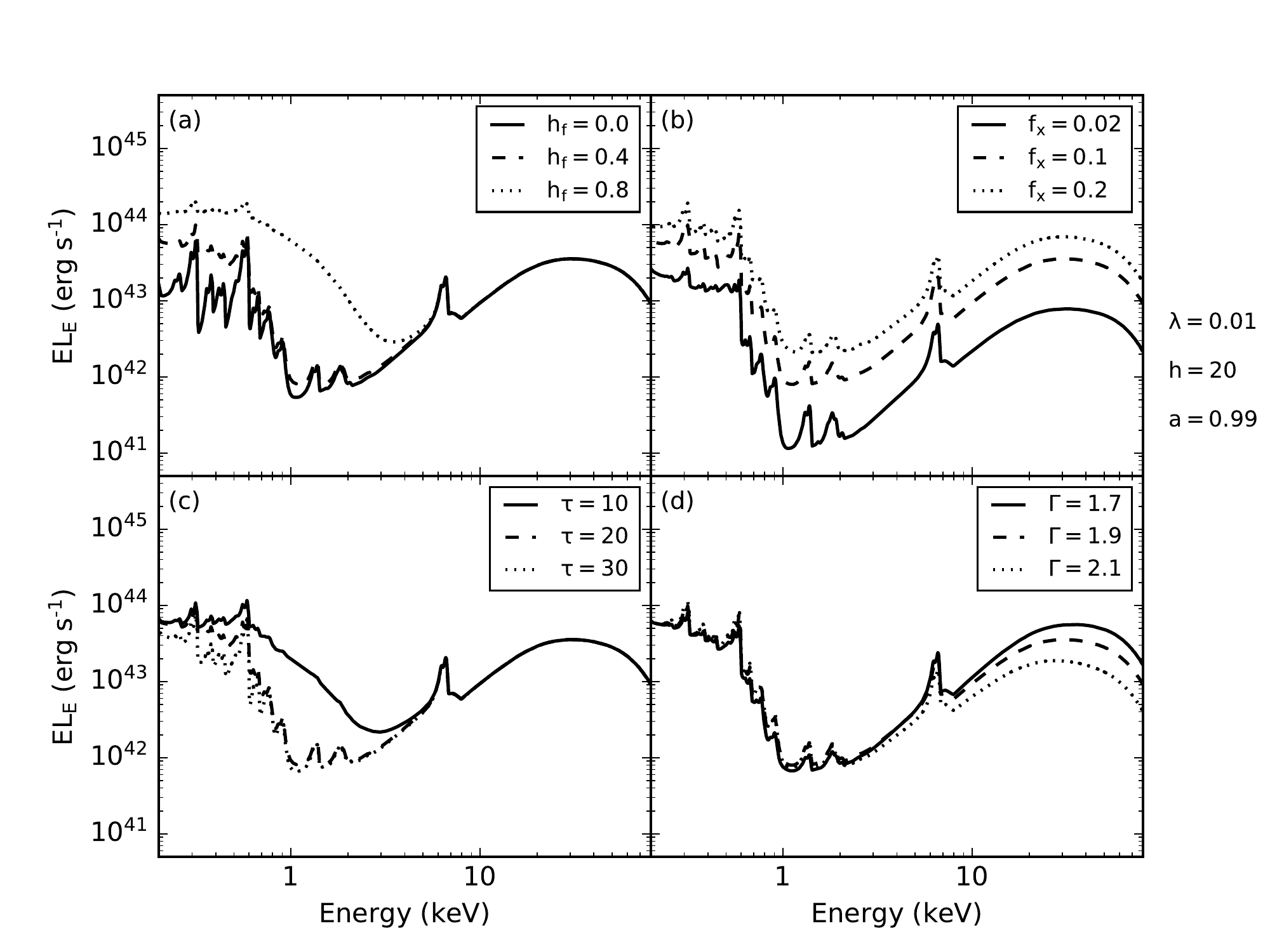}
\includegraphics[width=0.815\textwidth]{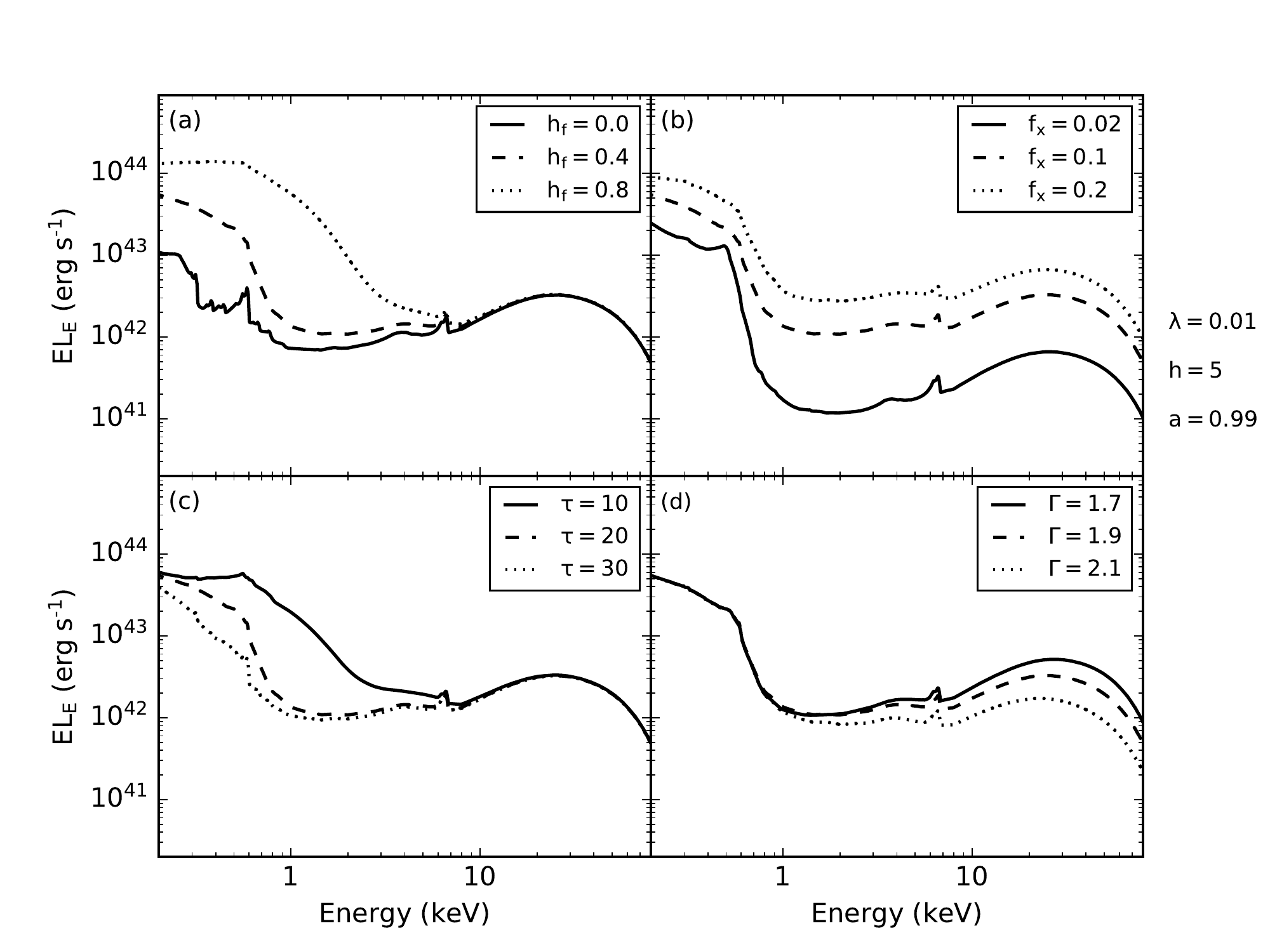}
\caption{(Top) As in Fig.~\ref{fig:l01h20a099}, but the spectra are taken from the grid with $\lambda=0.01$ $h=20$ and $a=0.99$. The smaller $\lambda$ leads to a denser and more weakly illuminated disc. The \rexcor\ spectra are therefore dominated by a low ionization reflector. However, a strong soft excess can still be produced with a sufficiently large $h_f$ or small $\tau$. (Bottom) In this case, $h=5$ and the strong light-bending from the low lamppost focuses the radiation onto the inner accretion disc. As a result of the steep ionization gradient on the disc, the \rexcor\ spectra become dominated by neutral reflection close to the black hole (Fig.~\ref{fig:rwarm}). The soft excess in this situation is then largely formed by the relativistic blurring of these reflection features \citep[e.g.,][]{crummy06}. However, as when $h=20$, additional warm coronal heating can smooth out and increase the strength of the soft excess.}
\label{fig:l001h20a099}
\end{figure*}

The upper-half of Figure~\ref{fig:l001h20a099} shows the \rexcor\ spectra from the $\lambda=0.01$, $h=20$, $a=0.99$ grid.
In contrast to Fig.~\ref{fig:l01h20a099}, where $\lambda=0.1$, these spectra frequently exhibit hallmarks of neutral reflection, including a relativistically broadened \fe\ line at 6.4~\kev\ and significant line emission below 1~\kev. The smaller $\lambda$ yields a denser and more weakly irradiated disc (Eqs.~\ref{eq:density} and~\ref{eq:Lx}) which reduces $r_{\mathrm{warm}}$ (Fig.~\ref{fig:rwarm}). The gas in the disc surface is much cooler in this scenario and produces a soft excess with many emission features. However, as seen in Fig.~\ref{fig:l001h20a099}(a,c), a sufficiently large $h_f$ in a small enough $\tau$ can raise the gas temperature to the point where a smooth soft excess is generated. In general, it is more challenging for a warm corona to be an important contributor to the soft excess when the gas is denser and can cool more efficiently \citep{ball20}.

When the lamppost height is reduced to $h=5$ (bottom half of Fig.~\ref{fig:l001h20a099}), light-bending causes the disc to be strongly illuminated at small radii, but the ionization parameter of the gas remains relatively low (Fig.~\ref{fig:rwarm}) resulting in \rexcor\ spectra dominated by highly relativistically blurred neutral reflection which significantly contributes to the soft excess \citep[e.g.][]{crummy06}. However, as seen in panels (a)-(c) of the figure, additional warm coronal heating is needed in this scenario to strengthen and smooth out the soft excess.

Overall, Figs.~\ref{fig:l01h20a099} and~\ref{fig:l001h20a099} show that the \rexcor\ grids provide spectra that produce a diverse range of spectral shapes, in particular for the soft excess. The combination of relativistically blurred reflection, an ionization gradient along the disc, and warm coronal heating lead to soft excesses with different strengths, slopes and smoothness. Applying a \rexcor\ grid to an AGN dataset will therefore allow an estimate of how the accretion power is distributed within the disc.

\section{examining the soft excess in agns with rexcor}
\label{sect:fits}
\begin{figure}
  \includegraphics[width=0.48\textwidth]{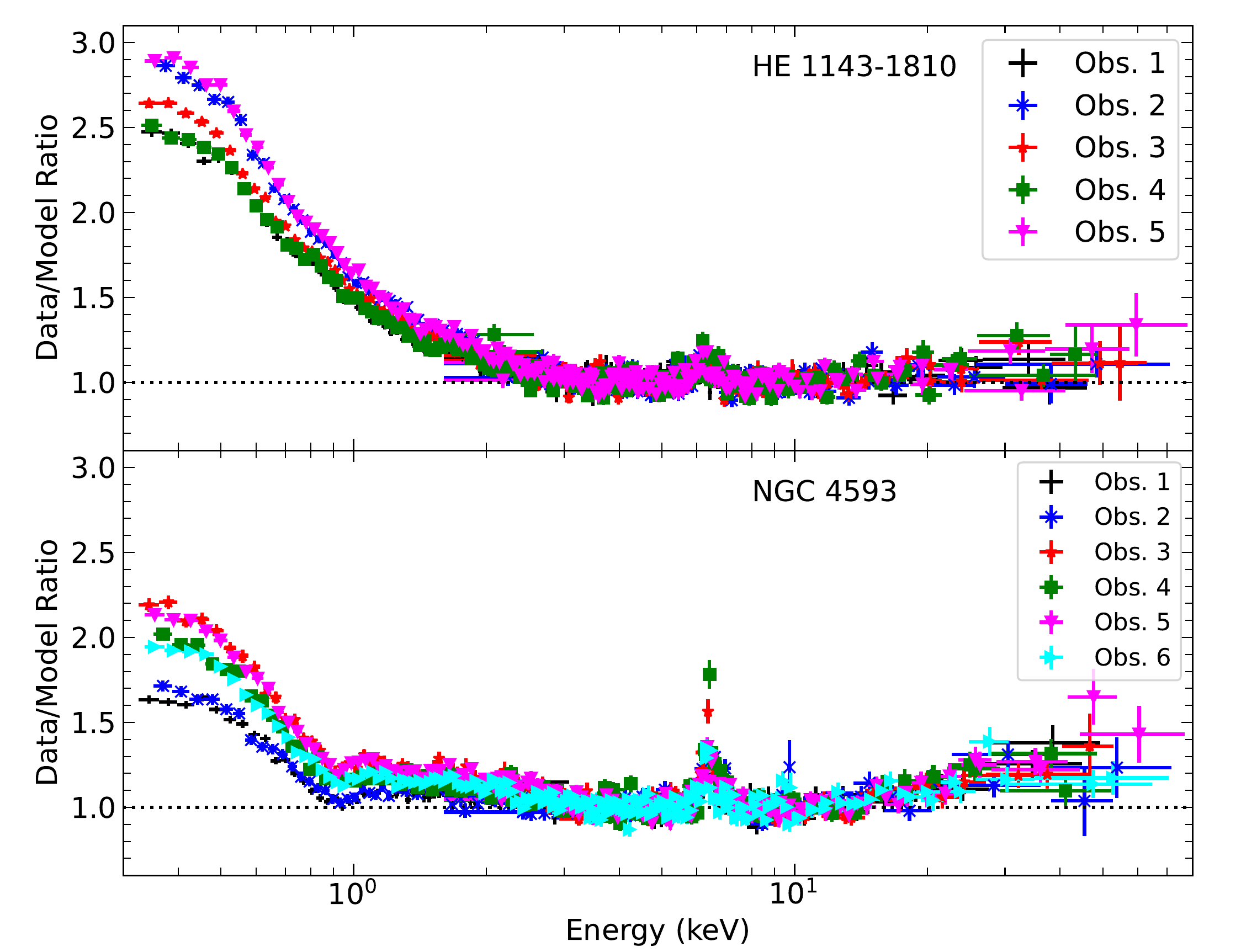}
    \caption{(Top) Data to model ratios found from the five joint
      \xmm/\nustar\ observations of \he\ when fit with a simple
      absorbed power-law in the energy range $3$--$6$~\kev\ and
      $7$--$20$~\kev\ and then extrapolated to the full energy band. A prominent \fe\ line, Compton reflection hump
      and strong soft excess are seen in all observations. (Bottom) As
      in the other panel, but now showing the results for the six
      observations of \ngc.} 
    \label{fig:absplawratios}
\end{figure}

In this section we illustrate the use of \rexcor\ in fitting AGN X-ray
data and demonstrate how the model can constrain interesting
properties of the energy flow in AGN accretion discs. We fit several joint
\xmm\ and \nustar\ spectra of two Seyfert 1 galaxies, \he\ and \ngc,
that exhibit both reflection features and a significant soft excess
(Fig.~\ref{fig:absplawratios}). The fits presented below are performed with XSPEC v.12.12.0g \citep{arn96}. Uncertainties on the fit parameters are the 90 per cent confident level for a single parameter (i.e., a $\Delta \chi^2=2.71$ criterion).

\subsection{\he}
\label{sub:he}
The Seyfert~1 \he\ ($z=0.0328$) was monitored by \xmm\ \citep{xmm01} and \nustar\ \citep{harr13} with five $20$~ks observations, each separated by two days. The spectral and timing analysis of this dataset was published by \citet{ursini20}, and we use the same five sets of EPIC-pn \citep{pn01} and \nustar\ spectra as described in that paper. As \rexcor\ is unable to make predictions at ultraviolet wavelengths, we do not analyze the data from \xmm\ Optical Monitor \citep{mason01}.

The Eddington ratio of \he\ is estimated to be $\lambda \sim 0.16$--$0.2$ \citep{ursini20}, but the object does not have a detectable broad \fe\ line and does not have a black hole spin measurement. Therefore, we choose the $\lambda=0.1$ and $a=0.99$ series of \rexcor\ grids to fit the data and test both the $h=20$ and $h=5$ cases\footnote{As expected, given the high ionization of the $\lambda=0.1$ models (Sect.~\ref{sub:grids}), fits with the $a=0.9$ grids result in similar $\chi^2$ values as the $a=0.99$ models. The parameters are also similar, with a 10\% rise in $\tau$ and a 50\% increase in $f_X$ (illustrating the degeneracy between $f_X$ and $a$; Sect.~\ref{sect:rexcor}).}. In addition to \rexcor, the spectral model includes a cutoff power-law (\textsc{zcutoffpl}) and neutral reflection from distant material (\textsc{xillver}; \citealt{garcia13}) which is needed to fit the narrow \fe\ line in the source. The cutoff energy in both of these components is fixed\footnote{Allowing the cutoff energy to vary did not lead to a significant improvement in the spectral fit.} at $100$~\kev\ \citep{ursini20}, and the $\Gamma$ is tied to the same value as in the \rexcor\ model. A solar iron abundance and an inclination angle of $30^{\circ}$ are assumed in the  \textsc{xillver} model. Neutral absorption due to a Galactic column density of $N_{\mathrm{H}}= 3.47\times 10^{20}$~cm$^{-2}$ \citep{kalberla05} is included using the \textsc{phabs} model.

The five observations are fit simultaneously with $\Gamma$, $h_f$, $f_X$, $\tau$, and the normalizations of the 3 spectral components (\textsc{zcutoffpl}, \rexcor, and \textsc{xillver}) free to vary for each observation. To simplify the procedure, we ignore the small constant offset between the two \nustar\ focal plane modules (FPMA and FPMB) in each observation (i.e., they are treated as one data group); however, a variable normalization constant is applied to the EPIC-pn data. Finally, as discussed by \citet{ursini20}, there is a small, energy-dependent offset in $\Gamma$ between \xmm\ and \nustar\ spectra of the same source. To correct for this, we follow \citet{ursini20} and apply a cross-calibration function to the \xmm\ data proportional to $E^{\Delta \Gamma}$, where $\Delta \Gamma = \Gamma^{\mathrm{XMM}} - \Gamma^{\mathrm{NuSTAR}}$ \citep{ingram17}. The results presented below report photon indices and fluxes using the \nustar\ data.

\begin{figure*}    
\includegraphics[width=0.49\textwidth]{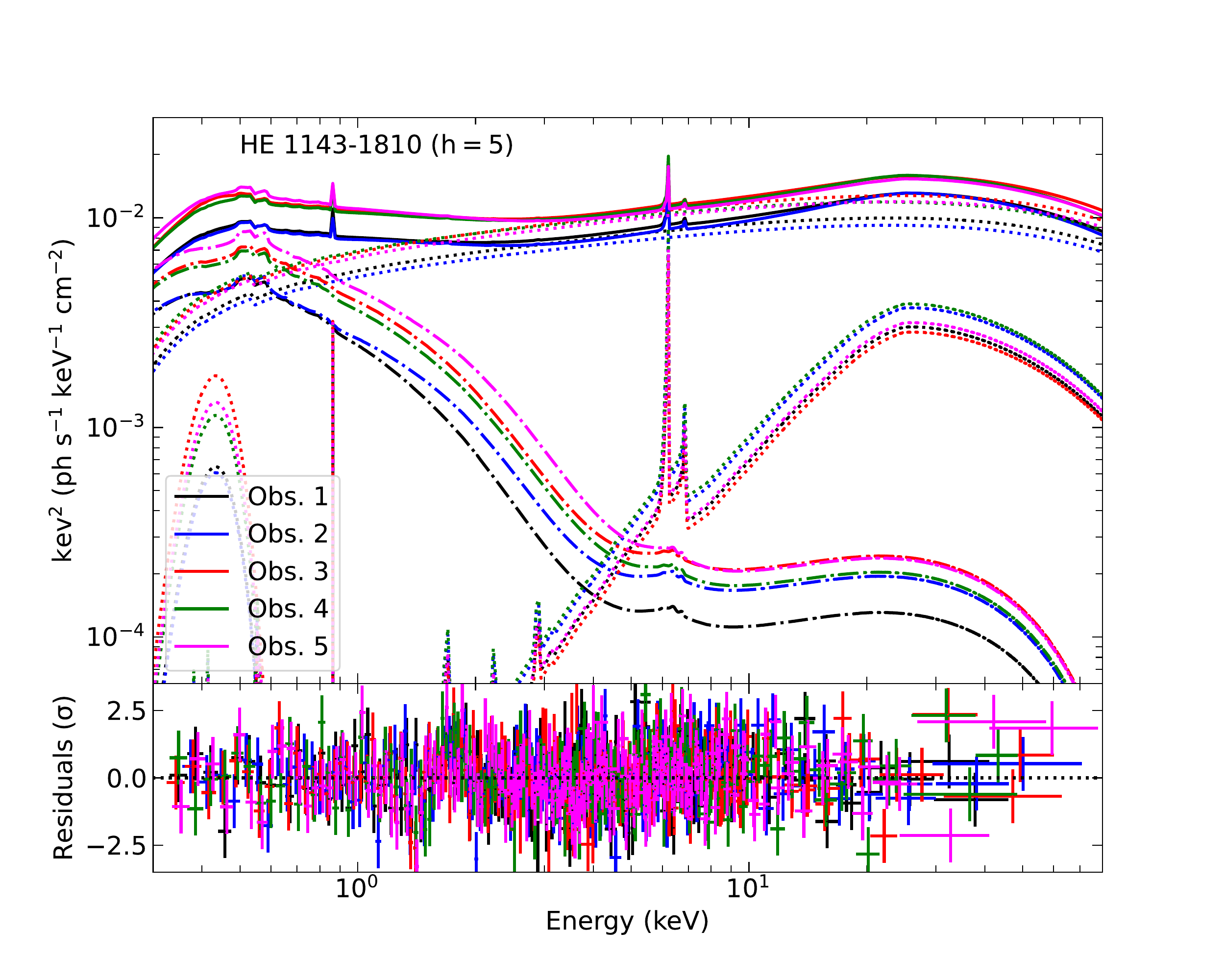}
\includegraphics[width=0.49\textwidth]{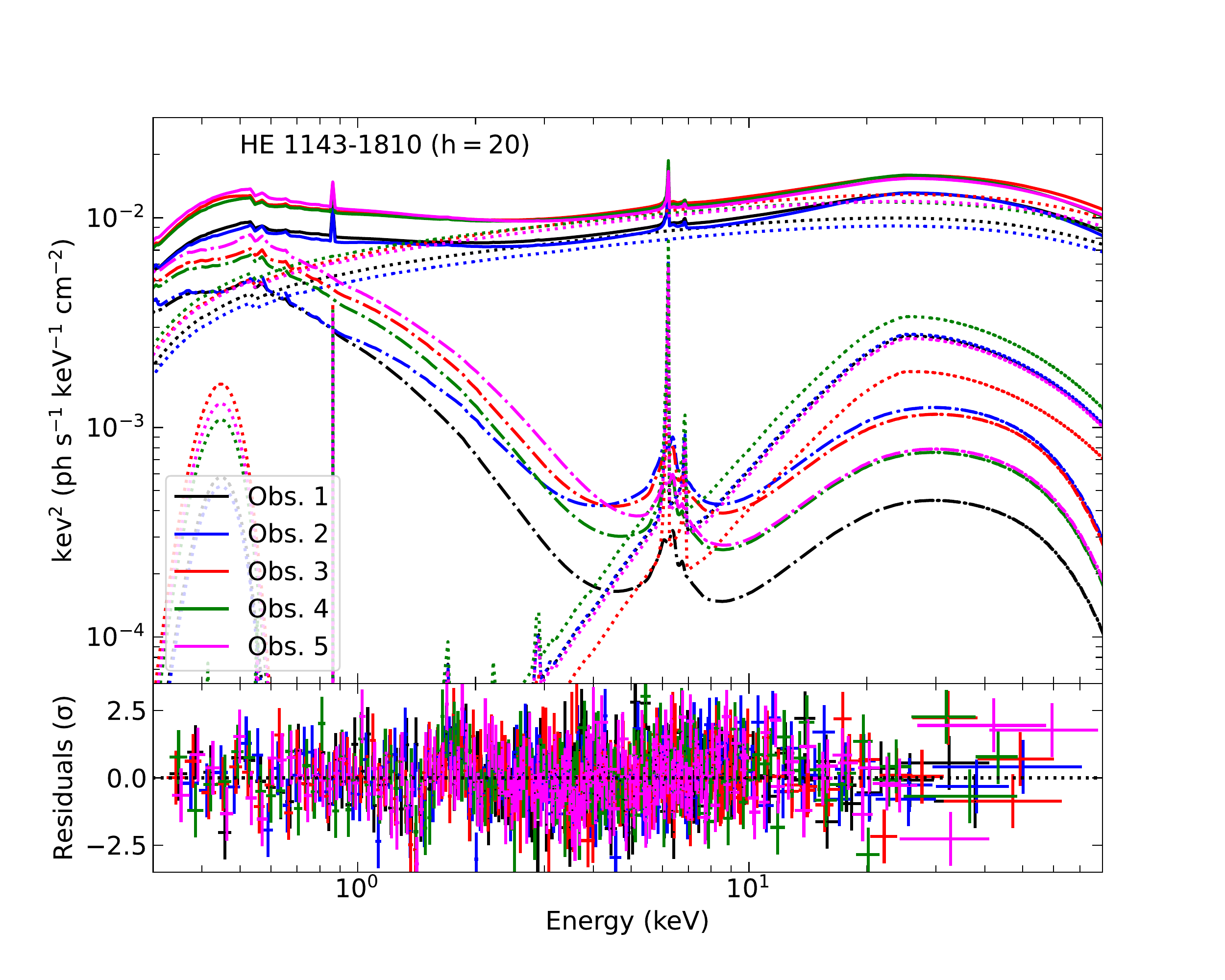}
    \caption{(Left) The upper panel plots the model components predicted from our $h=5$ fit (upper half of Table~\ref{table:hefit}) for each of the 5 observations of \he. The solid lines plot the total model, while the \rexcor\ components are shown as the dot-dashed lines. The dotted lines denote the remaining components of the model (the cutoff power-law, two Gaussian emission line, and $\textsc{xillver}$). This fit uses the \texttt{reXcor\_l01\_a099\_h5.fits} grid. The lower panel shows the residuals to the fit in units of $\sigma$. (Right) As in the other panel, but for the $h=20$ fit (lower half of Table~\ref{table:hefit}). This fit uses the \texttt{reXcor\_l01\_a099\_h20.fits} grid.}
    \label{fig:hefit}
\end{figure*}

Both the $h=5$ and $20$ \rexcor\ models yield very good fits to the \he\ data (with reduced $\chi^2 \la 1$), and the results are shown in Table~\ref{table:hefit} and Figure~\ref{fig:hefit}. 
\begin{table*}
    \centering
    \caption{Results from fitting five \xmm\ and \nustar\ observations of the Seyfert 1 galaxy \he\ with the following spectral model: \textsc{phabs}*(\textsc{zgauss1}+\textsc{zgauss2}+\textsc{zcutoffpl}+\rexcor+\textsc{xillver}). An energy dependent cross-calibration function proportional to $E^{\Delta \Gamma}$, where $\Delta \Gamma=\Gamma^{\mathrm{XMM}}-\Gamma^{\mathrm{NuSTAR}}$ is included in the model \citep{ursini20}. All five observations of the source were fit simultaneously between $0.3$ and $79$~\kev. These fits uses the $\lambda=0.1$, $a=0.99$, \rexcor\ grids, with the top half of the table showing the results from the $h=5$ grid ( $\chi^2/\mathrm{dof}=1934/1949$) and the lower half gives the results from the $h=20$ set of models ($\chi^2/\mathrm{dof}=1928/1949$). The two \textsc{zgauss} components likely arise from N~\textsc{vi} and Ne~\textsc{ix}, respectively. The latter line is unresolved (with $\sigma$ fixed at $0$~keV). The normalization of the lines ($K$) are in units of ph~s$^{-1}$~cm$^{-2}$, and their equivalent widths are listed as 'EW'. All fluxes ($F$) are tabulated in units of erg~s$^{-1}$~cm$^{-2}$. A 'p' in the errorbar indicates that the parameter pegged at the upper or lower limit of the grid. The photon-indices and fluxes are derived from the \nustar\ spectra.}  
    \label{table:hefit}
    \begin{tabular}{l|l|c|c|c|c|c|c}
 & & All Obs. & Obs. 1 & Obs. 2 & Obs. 3 & Obs. 4 & Obs. 5 \\ \hline
 & $\log F_{\mathrm{2-10\ keV}}$ & & $-10.65$ & $-10.67$ & $-10.55$ & $-10.56$ & $-10.57$ \\ \hline
 & & & & $h=5$ & & & \\ \hline
\textsc{zgauss1} & $E_1$ (keV) & $0.43^{+0.02}_{-0.01}$ & & & & \\
 & $\sigma_1$ (keV) & $0.06\pm 0.02$ & & & & \\
 & $K_1$ ($\times 10^{-3}$) & & $0.72\pm 0.03$ & $0.67\pm 0.31$ & $2.0\pm 0.6$ & $1.3\pm 0.5$ & $1.5^{+0.4}_{-0.5}$ \\
 & EW (eV) & & 11 & 11 & 22 & 15 & 15 \\
\textsc{zgauss2} & $E_2$ (keV) & $0.90^{+0.02}_{-0.01}$ & & & & \\
 & $K_2$ ($\times 10^{-5}$) & & $4.9^{+2.5}_{-2.3}$ & $3.3\pm 2.7$ & $5.7^{+2.9}_{-3.0}$ & $5.6^{+3.2}_{-1.6}$ & $5.6^{+3.0}_{-2.7}$ \\
 & EW (eV) & & 4 & 3 & 4 & 4 & 3 \\
\textsc{zcutoffpl} & $\Gamma$ & & $1.76^{+0.04}_{-0.03}$ & $1.76^{+0.01}_{-0.02}$ & $1.75^{+0.02}_{-0.03}$ & $1.77\pm 0.03$ & $1.75\pm 0.03$ \\
 & $\Delta \Gamma$ & & $-0.042$ & $-0.065$ & $-0.029$ & $-0.045$ & $-0.052$ \\
\textsc{reXcor} & $f_X$ & & $0.045^{+0.015}_{-0.006}$ & $0.05\pm 0.01$ & $0.060\pm 0.007$ & $0.052\pm 0.012$ & $0.05\pm 0.01$ \\
 & $h_f$ & & $0.45\pm 0.09$ & $0.39\pm 0.05$ & $0.45^{+0.1}_{-0.07}$ & $0.45^{+0.14}_{-0.08}$ & $0.45^{+0.01}_{-0.04}$ \\
 & $\tau$ & & $11.6^{+1.4}_{-1.0}$ & $10.2^{+0.8}_{-0.2p}$ & $11.3^{+0.6}_{-0.8}$ & $11.1^{+1.2}_{-0.8}$ & $10.4^{+0.6}_{-0.4p}$ \\
 & & & & & & & \\
 & $\log F_{\mathrm{0.3-10\ keV}}^{\mathrm{PL}}$ & & $-10.42\pm 0.01$ & $-10.45\pm 0.01$ & $-10.33\pm 0.01$ & $-10.33\pm 0.01$ & $-10.35^{+0.01}_{-0.003}$ \\ 
 & $\log F_{\mathrm{0.3-10\ keV}}^{\mathrm{reXcor}}$ & & $-10.90^{+0.06}_{-0.05}$ & $-10.88^{+0.05}_{-0.06}$ & $-10.72^{+0.06}_{-0.05}$ & $-10.75^{+0.05}_{-0.06}$ & $-10.65^{+0.04}_{-0.03}$ \\
 & $\log F_{\mathrm{0.3-10\ keV}}^{\mathrm{xillver}}$ & & $-12.08^{+0.09}_{-0.10}$ & $-11.99^{+0.05}_{-0.08}$ & $-12.12^{+0.09}_{-0.14}$ & $-11.96^{+0.09}_{-0.11}$ & $-12.07\pm 0.10$\\
 & $\chi^2/\mathrm{dof}$ & $1934/1949$ & & & & \\
    \hline
    & & & & $h=20$ & & & \\ \hline
\textsc{zgauss1} & $E_1$ (keV) & $0.44^{+0.02}_{-0.03}$ & & & & \\
 & $\sigma_1$ (keV) & $0.06\pm 0.02$ & & & & \\
 & $K_1$ ($\times 10^{-3}$) & & $0.62^{+0.04}_{-0.03}$ & $0.56^{+0.04}_{-0.03}$ & $1.7^{+1}_{-0.5}$ & $1.2^{+0.7}_{-0.4}$ & $1.4^{+0.8}_{-0.6}$ \\
 & EW (eV) & & 10 & 9 & 21 & 15 & 16 \\
\textsc{zgauss2} & $E_2$ (keV) & $0.89^{+0.01}_{-0.02}$ & & & & \\
 & $K_2$ ($\times 10^{-5}$) & & $5.7^{+2.7}_{-2.4}$ & $5.7^{+3.1}_{-3.3}$ & $7.0\pm 3.7$ & $6.6^{+4.2}_{-3.5}$ & $6.4^{+3.5}_{-3.4}$ \\
 & EW (eV) & & 5 & 5 & 4 & 4 & 4 \\
\textsc{zcutoffpl} & $\Gamma$ & & $1.76^{+0.04}_{-0.03}$ & $1.76\pm 0.02$ & $1.73^{+0.03}_{-0.02}$ & $1.77^{+0.03}_{-0.04}$ & $1.75^{+0.03}_{-0.02}$ \\
 & $\Delta \Gamma$ & & $-0.040$ & $-0.056$ & $-0.024$ & $-0.042$ & $-0.047$ \\
\textsc{reXcor} & $f_X$ & & $0.028^{+0.025}_{-0.008p}$ & $0.051^{+0.011}_{-0.016}$ & $0.050^{+0.021}_{-0.020}$ & $0.038^{+0.024}_{-0.018p}$ & $0.033^{+0.018}_{-0.013p}$ \\
 & $h_f$ & & $0.50^{+0.06}_{-0.08}$ & $0.36^{+0.14}_{-0.07}$ & $0.49^{+0.08}_{-0.13}$ & $0.50^{+0.08}_{-0.14}$ & $0.52^{+0.04}_{-0.09}$ \\
 & $\tau$ & & $12.1^{+1.8}_{-0.8}$ & $10.4^{1.5}_{-0.4p}$ & $11.9^{+1.5}_{-1.1}$ & $11.8^{+1.3}_{-0.7}$ & $11.1^{+0.6}_{-0.5}$ \\
 & & & & & & & \\
 & $\log F_{\mathrm{0.3-10\ keV}}^{\mathrm{PL}}$ & & $-10.42\pm 0.01$ & $-10.46\pm 0.01$ & $-10.34^{+0.01}_{-0.02}$ & $-10.33\pm 0.02$ & $-10.36\pm 0.01$ \\ 
 & $\log F_{\mathrm{0.3-10\ keV}}^{\mathrm{reXcor}}$ & & $-10.90\pm 0.06$ & $-10.86\pm 0.05$ & $-10.71\pm 0.06$ & $-10.76^{+0.08}_{-0.06}$ & $-10.66^{+0.06}_{-0.05}$ \\
 & $\log F_{\mathrm{0.3-10\ keV}}^{\mathrm{xillver}}$ & & $-12.13^{+0.13}_{-0.17}$ & $-12.12^{+0.11}_{-0.15}$ & $-12.32^{+0.19}_{-0.32}$ & $-12.02^{+0.12}_{-0.18}$ & $-12.15^{+0.13}_{-0.18}$\\
 & $\chi^2/\mathrm{dof}$ & $1928/1949$ & & & & \\
    \hline
    \end{tabular}
\end{table*}
We first focus on the best fit values for the three \rexcor\ parameters that describe the distribution of accretion energy between the warm and hot coronas (i.e., $h_f$, $f_X$ and $\tau$). The values of these parameters do not significantly differ between the $h=5$ and $h=20$ models. Thus, while we are unable to place a constraint on the lamppost height, the properties of the soft excess in \he\ allow a robust measurement of the warm corona properties. Notably, we find that $h_f \neq 0$ in all five observations which indicates that warm corona heating is required to account for the soft excess in \he. The optical depth of the warm corona is consistently low and varies only from $\approx 10$ to $\approx 12$. These values are slightly less than those inferred by \citet{ursini20} (where $\tau \approx 17.5$) using a model which describes the soft excess with only a Comptonization spectrum. As seen in Fig.~\ref{fig:l01h20a099}, a low value of $\tau$ maximizes the heating effects for a given $h_f$, and will lead to a stronger and smoother soft exess, as observed in \he. The values of $f_X$ indicate that $\la 6$\% of the accretion energy is dissipated in the lamppost corona, consistent with the observed bolometric corrections in AGNs at these Eddington ratios \citep[e.g.,][]{duras20}.

Interestingly, each fit requires the addition of 2 Gaussian emission line components, a broadened one ($\sigma=0.06$~\kev) at $\approx 0.43$~\kev\ and an unresolved ($\sigma=0$~\kev) at $\approx 0.9$~\kev. The lines are highly statistically significant (with F-test probabilities $\ll 10^{-7}$), but have equivalent widths (EWs) of $3$--$22$~eV. The energy of the broadened $0.43$~\kev\ line is consistent with the  N~\textsc{vi} triplet, and its EW increases from $11$~eV to $22$~eV as the source brightens. This fact, combined with its width, suggests that this line is responding to the changing ionization state of the inner accretion disc. The soft excess produced by \rexcor\ is comprised of blurred ionized reflection superimposed on a Comptonized continuum enhanced by the warm corona. The \rexcor\ spectra include emission from N~\textsc{vi}, but appears to underestimate its strength in \he\ by $\approx 10$\%. This mismatch is likely  a result of fact that transitions in He-like ions such as N~\textsc{vi} are very sensitive to the temperature, density and optical depth of a plasma \citep{pdg10}, and these conditions are not correctly described by the \rexcor\ models for \he. In contrast, the EW of the narrow $0.9$~\kev\ line is roughly constant across all observations, indicating that it originates from an unchanging ionized zone at some distance from the black hole. The line energy is consistent with arising from Ne~\textsc{ix}, which is not included in the \rexcor\ model, and therefore this line had to be included as a separate component in the fits. The presence of these emission lines is evidence that the soft excess in \he\ is comprised, in part, from photoionized emission across a range of ionization states. Therefore, given the complexity of emission features in this energy range, combined with the available number of lines predicted by \rexcor, we expect that the need to add additional Gaussian components will be common when applying \rexcor\ to AGN data.

\begin{figure}    
\includegraphics[width=0.48\textwidth]{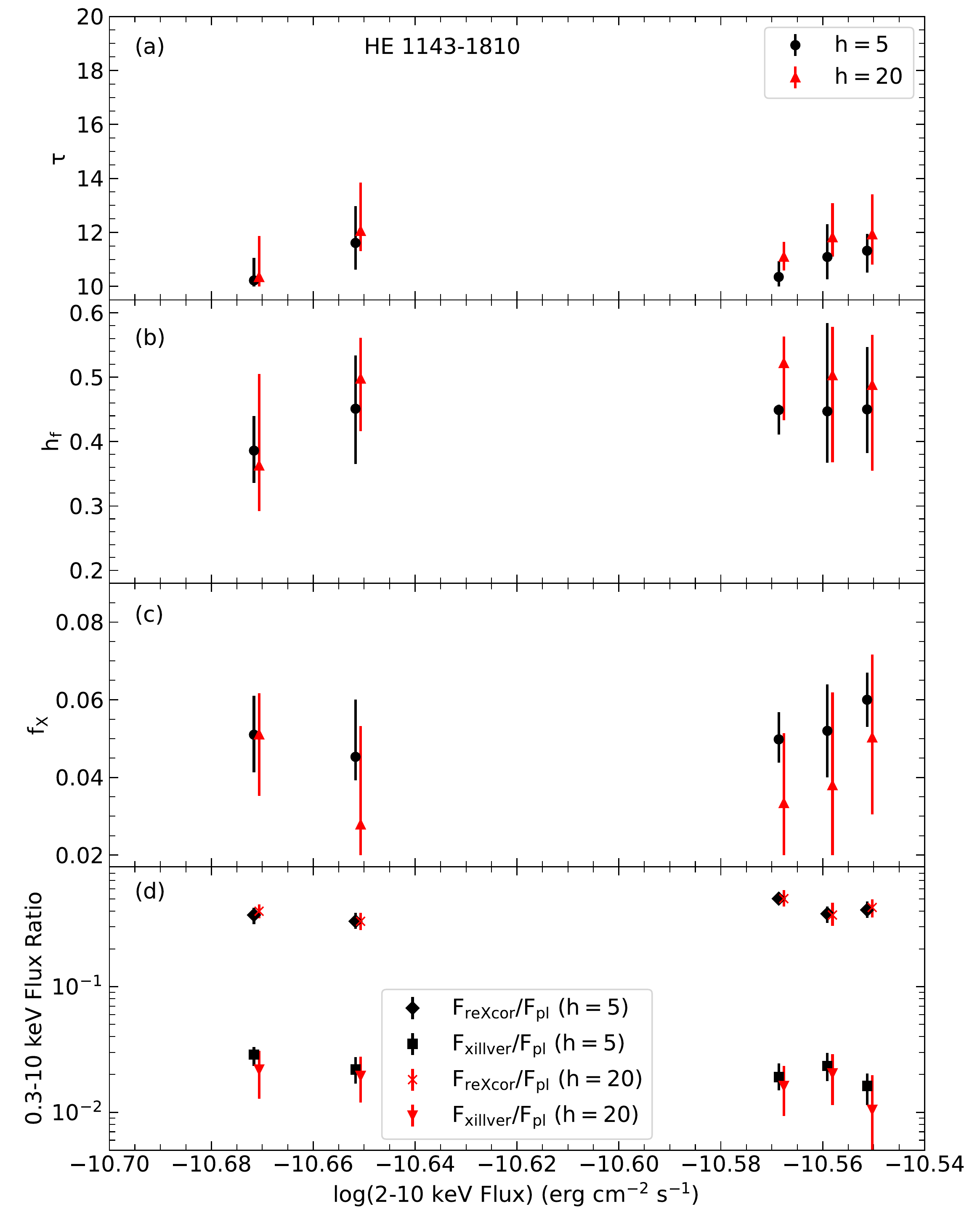}
    \caption{Panels (a) (b) and (c) plot the best-fit values of $\tau$, $h_f$ and $f_X$ from the \rexcor\ fits to \he\ (Table~\ref{table:hefit} and Fig.~\ref{fig:hefit}) against the observed $2$--$10$~\kev\ flux. The results from the $h=5$ model are shown as the black circles, while the red triangles plot the parameters from the $h=20$ model. The lower panel shows that the fraction of the flux described by both the \rexcor\ and \textsc{xillver} models remains relatively unchanged in each of the five observations for both the $h=5$ and $h=20$ models.}
    \label{fig:heparams}
\end{figure}

As the five observations of \he\ span a factor of $\approx 2$ in flux, it is interesting to consider how the \rexcor\ parameters change as the source changed in brightness. Panels (a)--(c) of Figure~\ref{fig:heparams} plot $\tau$, $h_f$ and $f_X$ as a function of the observed $2$--$10$~\kev\ flux of \he\ for both the $h=5$ (black circles) and $h=20$ (red triangles) fits. 
There is no evidence for a correlation (i.e., a linear fit returns a slope consistent with zero) between the these parameters and the observed flux in either scenario. 
Although an increase in the optical depth with flux is a common prediction of accretion disc models \citep[e.g.,][]{sz94,jiang19}, the exact dependence remains uncertain and it is not obvious if such a relationship extends to the warm corona. These results are consistent with the fits of \citet{ursini20} using a Comptonization model. It is possible that observations spanning a larger range in flux are necessary to detect variations in these parameters.

The normalization of the \rexcor\ models is determined by many quantities, including the distance to the source, the area of the disc emission region, the inclination angle of the disc, the black hole mass and spin, and the geometry of the X-ray source. As it is impractical to unravel all these effects to interpret the normalization returned by the fits, we consider the ratio of the \rexcor\ and power-law fluxes. The scenario used to calculate \rexcor\ spectra would predict a ratio close to unity, as the flux from the disc is produced by reprocessing the luminosity of the lamppost (Sect.~\ref{sect:model}). However, this ratio could be significantly altered by effects not included in our model, such as a non-stationary X-ray emitting corona \citep[e.g.,][]{belo99}, a truncated accretion disc \citep[e.g.,][]{kd18}, or a non-flat disc surface that could enhance the overall reflection strength \citep[e.g.][]{fab02}. Fig.~\ref{fig:heparams}(d) shows that the total \rexcor\ flux (in the $0.3$--$10$~keV band) in the \he\ fits is a factor of $\approx 0.4$ of the power-law flux, independent of the assumed coronal height. This flux ratio could be explained by a moderately outflowing corona \citep{belo99} or a truncated disc. Finally, we find that the flux of the \textsc{xillver} model is $\approx 0.02\times$ less than the power-law flux between $0.3$--$10$~\kev\ because of the very small albedo of dense, neutral gas in this energy range.


\subsection{\ngc}
\label{sub:ngc}
As a second example, we apply the \rexcor\ model to \ngc\ ($z=0.0831$), a Seyfert~1 with $\lambda \approx 0.04$ \citep{vf09}. Similar to \he, \ngc\ has five coordinated $20$~ks \xmm\ and \nustar\ observations that span a factor of $\approx 2$ in flux \citep{ursini16,middei19}. The first observation caught the source rapidly declining in flux, and so \citet{ursini16} split this observation into two in order to isolate the low count-rate region which exhibits a hard spectral shape. The resulting six EPIC-pn and \nustar\ spectra are analyzed here.

We apply a similar spectral model to \ngc\ as with \he: neutral absorption with \textsc{phabs} ($N_H=1.89\times 10^{22}$~cm$^{-2}$; \citealt{kalberla05}), a neutral \textsc{xillver} model to account for distant reflection, a cutoff power-law, and \rexcor. Given the lower Eddington ratio in this source, we use the $\lambda=0.01$ \rexcor\ grids. Although the source shows evidence for a broadened \fe\ line \citep{brenn07,ursini16}, a spin estimate does not exist. Therefore, we begin our analysis with the $a=0.99$ \rexcor\ grids, but also test the result with the $a=0.9$ grids. In addition, \ngc\ has been previously fit with two warm absorbers \citep{brenn07,ursini16} and a photoionized emitter \citep{ursini16} to account for the observed soft X-ray spectral complexity. Therefore, we include a warm absorber table model calculated with XSTAR \citep{walton13} in our fit, but we find that a distinct photoionized plasma emission model is not required. In contrast with the \he\ fits, the cutoff energy of the \textsc{zcutoffpl} model (which is tied to the corresponding parameter in \textsc{xillver}) is allowed to vary in each observation as \citet{ursini16} found that the cutoff energy is large and variable in these observations. Similarly, the iron abundance of the \textsc{xillver} model is allowed to vary, although the value is the same for all observations. 

\begin{figure}    
\includegraphics[width=0.5\textwidth]{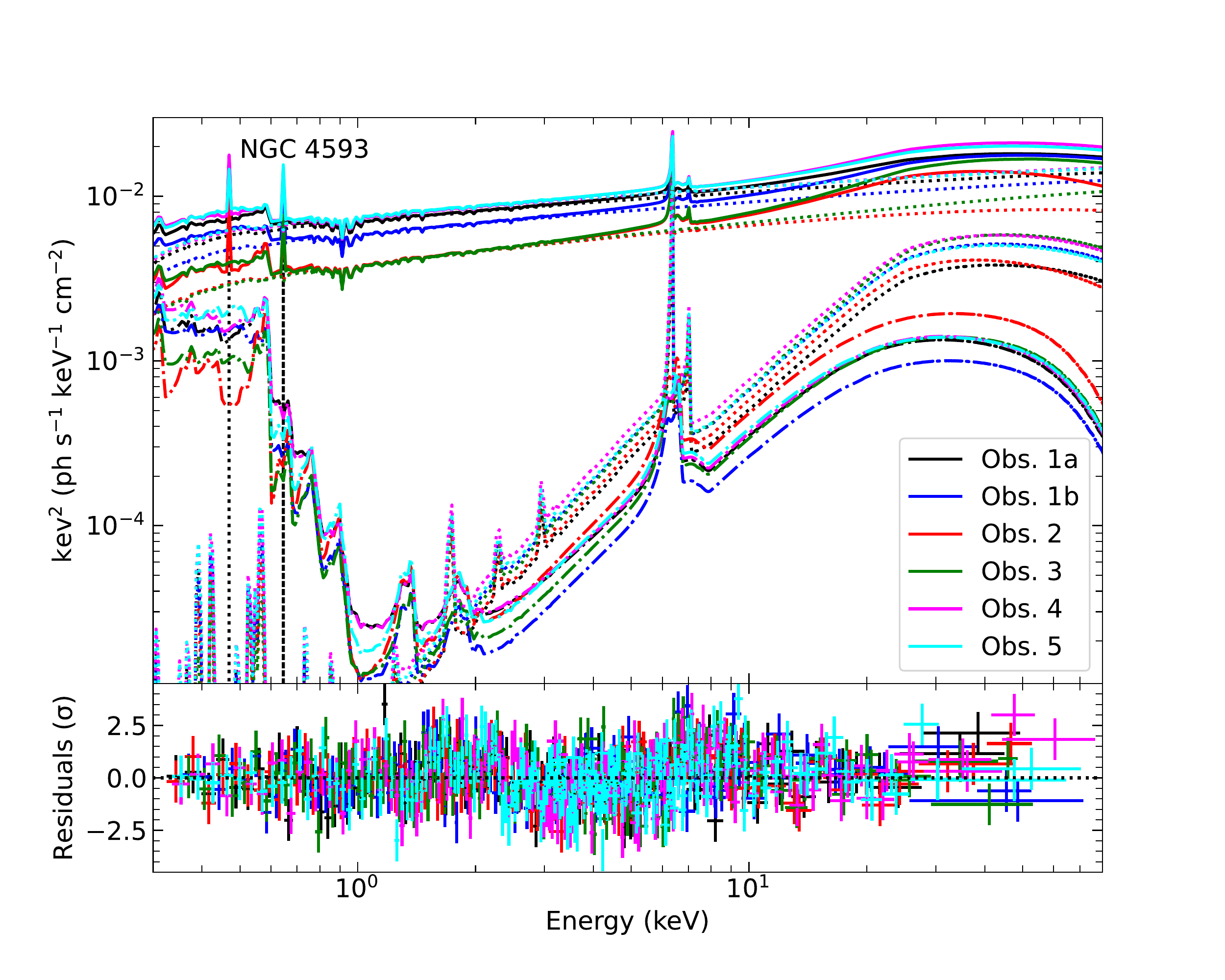}
    \caption{The upper panel plots the model components predicted from our best fit (Table~\ref{table:ngcfit}) for each of the 6 observations of \ngc. The solid lines plot the total model, while the \rexcor\ components are shown as the dot-dashed lines. The dotted lines denote the remaining components of the model (the cutoff power-law, two Gaussian emission lines, and $\textsc{xillver}$). This model uses the \texttt{reXcor\_l001\_a099\_h20.fits} grid. The lower panel shows the residuals to the fit in units of $\sigma$.  }
    \label{fig:ngcfit}
\end{figure}

The fit procedure for \ngc\ is the same as for \he, including the use of the $\Delta \Gamma$ cross-calibration function to account for the small offset in the \xmm\ and \nustar\ photon indices. As before, we report fluxes and $\Gamma$ from the \nustar\ data. The best fit model (with $\chi^2/\mathrm{dof}=2450/2271$) is obtained with the \texttt{reXcor\_l001\_a099\_h20.fits} grid and is shown in Figure~\ref{fig:ngcfit} with the results tabulated in Table~\ref{table:ngcfit}.
\begin{table*}
    \centering
    \caption{Results from fitting five \xmm\ and \nustar\ observations of the Seyfert 1 galaxy \ngc\ with the following spectral model: \textsc{phabs}*\textsc{WA}*(\textsc{zgauss1}+\textsc{zgauss2}+\textsc{zcutoffpl}+\rexcor+\textsc{xillver}), where \textsc{WA} denotes a warm absorber table model \citep{walton13}.  An energy dependent cross-calibration function proportional to $E^{\Delta \Gamma}$, where $\Delta \Gamma=\Gamma^{\mathrm{XMM}}-\Gamma^{\mathrm{NuSTAR}}$ is included in the model \citep{ursini20}. The first observation is split in two \citep[e.g.,][]{ursini16} and all six spectra are fit simultaneously between $0.3$ and $79$~\kev. This fit uses the $\lambda=0.01$, $a=0.99$, $h=20$ \rexcor\ grid. The two \textsc{zgauss} components have $\sigma=0$~keV with normalizations ($K$) in units of ph~s$^{-1}$~cm$^{-2}$. All fluxes ($F$) are tabulated in units of erg~s$^{-1}$~cm$^{-2}$. A 'p' in the errorbar indicates that the parameter pegged at the upper or lower limit of the grid. The photon-indices and fluxes are derived from the \nustar\ spectra.}  
    \label{table:ngcfit}
    \begin{tabular}{l|l|c|c|c|c|c|c|c}
    \hline
 & & All Obs. & Obs. 1a & Obs. 1b & Obs. 2 & Obs. 3 & Obs. 4 & Obs. 5 \\ \hline
 \textsc{WA} & $N_{\mathrm{H}}$ ($\times 10^{21}$~cm$^{-2}$) & $2.75^{+0.38}_{-0.39}$ & & & & & \\
 & $\log \xi$ (erg~s~cm$^{-1}$) & $2.33^{+0.06}_{-0.05}$ & & & & & & \\
\textsc{xillver} & $A_{\mathrm{Fe}}$ & $2.7^{+0.5}_{-0.2}$ & & & & & & \\
\textsc{zgauss1} & $E_1$ (keV) & $0.47\pm 0.01$ & & & & & & \\
 & $K_1$ ($\times 10^{-4}$) &  & $3.6^{+2.1}_{-2.6}$ & $0.02{+3}_{-0.2p}$ & $2^{+1}_{-1.9}$ & $0.01^{+2}_{-0.01p}$ & $3.4^{+1.7}_{-2.5}$ & $2.2^{+3.0}_{-1.1}$ \\
  & EW (eV) & & 10 & 0 & 11 & 0 & 9 & 5 \\
\textsc{zgauss2} & $E_2$ (keV) & $0.65\pm 0.01$ & & & & & & \\
 & $K_2$ ($\times 10^{-4}$) & & $1.7^{+0.8}_{-0.7}$ & $1.1^{+0.5}_{-0.7}$ & $0.7^{+0.2}_{-0.4}$ & $1.2\pm 0.4$ & $2.1^{+0.7}_{-0.8}$ & $2.3^{+0.3}_{-1}$ \\
  & EW (eV) & & 9 & 6 & 7 & 12 & 10 & 11 \\
\textsc{zcutoffpl} & $\Gamma$ & & $1.84\pm 0.04$ & $1.82^{+0.03}_{-0.06}$ &  $1.76^{+0.05}_{-0.01}$ & $1.75^{+0.04}_{-0.03}$ & $1.84^{+0.02}_{-0.03}$ & $1.84\pm 0.03$ \\
 & $E_{\mathrm{cut}}$ (keV) & & $>240$ & $>210$ & $>140$ & $>240$ & $>700$ & $>410$ \\
 & $\Delta \Gamma$ & & $-0.025$ & $-0.065$ & $-0.065$ & $-0.045$ & $-0.021$ & $-0.016$ \\
\textsc{reXcor} & $f_X$ & & $0.060^{+0.14p}_{-0.040p}$ & $0.02^{+0.066}_{-0p}$ & $0.02^{+0.067}_{-0p}$ & $0.029^{+0.152}_{-0.008}$ & $0.060^{+0.098}_{-0.040p}$ & $0.021^{+0.1}_{-0.001p}$ \\
 & $h_f$ & & $0.50^{+0.13}_{-0.19}$ & $0.50^{+0.19}_{-0.23}$ & $0.38^{+0.10}_{-0.23}$ & $0.47^{+0.21}_{-0.38}$ & $0.60^{+0.11}_{-0.15}$ & $0.45^{+0.23}_{-0.21}$ \\
 & $\tau$ & & $22.2^{+7.8p}_{-8.2}$ & $24.5^{+5.5p}_{-10.8}$ & $30^{+0p}_{-14.3}$ & $27.8^{+2.2p}_{-17.6}$ & $28.1^{+1.9p}_{-9.8}$ & $23.4^{+6.6p}_{-4.8}$ \\\\
 & $\log F_{\mathrm{0.3-10\ keV}}^{\mathrm{PL}}$ & &  $-10.34^{+0.02}_{-0.01}$ & $-10.41\pm 0.01$ & $-10.58\pm 0.01$ & $-10.58\pm 0.01$ & $-10.31\pm 0.01$ & $-10.31\pm 0.01$ \\
 & $\log F_{\mathrm{0.3-10\ keV}}^{\mathrm{reXcor}}$ & & $-11.52^{+0.07}_{-0.06}$ & $-11.59^{+0.07}_{-0.08}$ & $-11.70^{+0.06}_{-0.05}$ & $-11.69^{+0.05}_{-0.08}$ & $-11.46^{+0.04}_{-0.05}$ & $-11.47^{+0.04}_{-0.06}$ \\
 & $\log F_{\mathrm{0.3-10\ keV}}^{\mathrm{xillver}}$ & & $-12.07^{+0.10}_{-0.16}$ & $-11.96^{+0.09}_{-0.14}$ & $-12.02\pm 0.08$ & $-11.96\pm 0.10$ & $-11.89\pm 0.08$ & $-11.95^{+0.05}_{-0.10}$ \\
 & $\log F_{\mathrm{2-10\ keV}}$ & & $-10.6$ & $-10.66$ & $-10.8$ & $-10.79$ & $-10.5644$ & $-10.5645$ \\
 & $\chi^2/\mathrm{dof}$ & $2450/2271$ & & & & & \\
   \hline
    \end{tabular}
\end{table*}
The $h=5$ grid yields a worse fit ($\chi^2/\mathrm{dof}=2472/2271$), with similar average values of $f_X$ and $h_f$ as the $h=20$ model. The average value of $\tau$ drops from $26$ with the $h=20$ grid to $17$ with the $h=5$ models. Overall, it appears that a higher coronal height is preferred in \ngc. However, fits using $a=0.9$ grids do not appreciably change the goodness of fit compared to the $a=0.99$ grids, so we are unable to provide a constraint on the black hole spin. Interestingly, the best-fit model requires only a single, moderately ionized warm absorber. When a second warm absorber was added to the model, its column density was driven to very low values, effectively eliminating any impact on the spectrum.  

The \rexcor\ spectra in the \ngc\ fit show many features of reflection from a weakly ionized accretion disc. This is a result of the $\lambda=0.01$, $h=20$ grid which naturally produces spectra with a relatively small ionized zone (Fig.~\ref{fig:rwarm}). Nevertheless, Table~\ref{table:ngcfit} shows that $h_f\sim 0.5$ in all observations which indicates that warm corona heating is necessary to satisfactorily account for the observed soft excess. Fig.~\ref{fig:l001h20a099}(a) shows that heating at this level typically raises the soft emission at energies $\la 1$~\kev. The warm corona depth in \ngc\ are typically quite large with $\tau \ga 20$ and hitting the upper-limit of $30$ in each observation. A large value of $\tau$ for the warm corona is consistent with the fits of \citet{middei19} using a Comptonization model. As discussed in Sect.~\ref{sect:rexcor}, a large value of $\tau$ spreads out the heat from a given $h_f$ which will limit the temperature increase in the corona. The average value of $\tau$ found for \ngc\ is $26$, much larger than the mean $\tau$ of $11$ measured in the higher $\lambda$ \he\ with a stronger and smoother soft excess. Finally, we find that the value of $f_X$ in \ngc\ is frequently consistent with the lower limit of the grids, $f_X \sim 0.02$, indicating that the lamp-post is receiving a small fraction of the accretion power.

Fig.~\ref{fig:ngcfit} shows that the soft excess predicted by the \rexcor\ model is imprinted with several broadened emission features arising from reflection off the disc. This structured soft excess appears to eliminate the need for a second warm absorber component, as well as the photoionized emitter, that was used in earlier models \citep{brenn07,ursini16}. The need for these additional models likely was a result of assuming a smooth spectral model for the soft excess (such as a Comptonized model, or a bremsstrahlung spectrum). However, the \rexcor\ model naturally includes both photoionized and Comptonized emission, in addition to bremsstrahlung, and therefore yields a more straightforward model of the spectrum.

As in \he, two Gaussian emission lines are needed in the final model, but, in this case, both lines are narrow. The lower-energy line (at $0.46$~keV) is weak (with a normalization consistent with zero in Obs. 1b and 3), and could be associated with a blend of the Ly$\beta$ line and the radiative recombination continuum from C~\textsc{vi}. The $0.65$~\kev\ line is likely O~\textsc{viii} Ly$\alpha$, and is also weak with an EW$\sim 10$~eV. These lines will originate in distant ionized gas not connected to the \rexcor\ model.

\begin{figure}    
\includegraphics[width=0.48\textwidth]{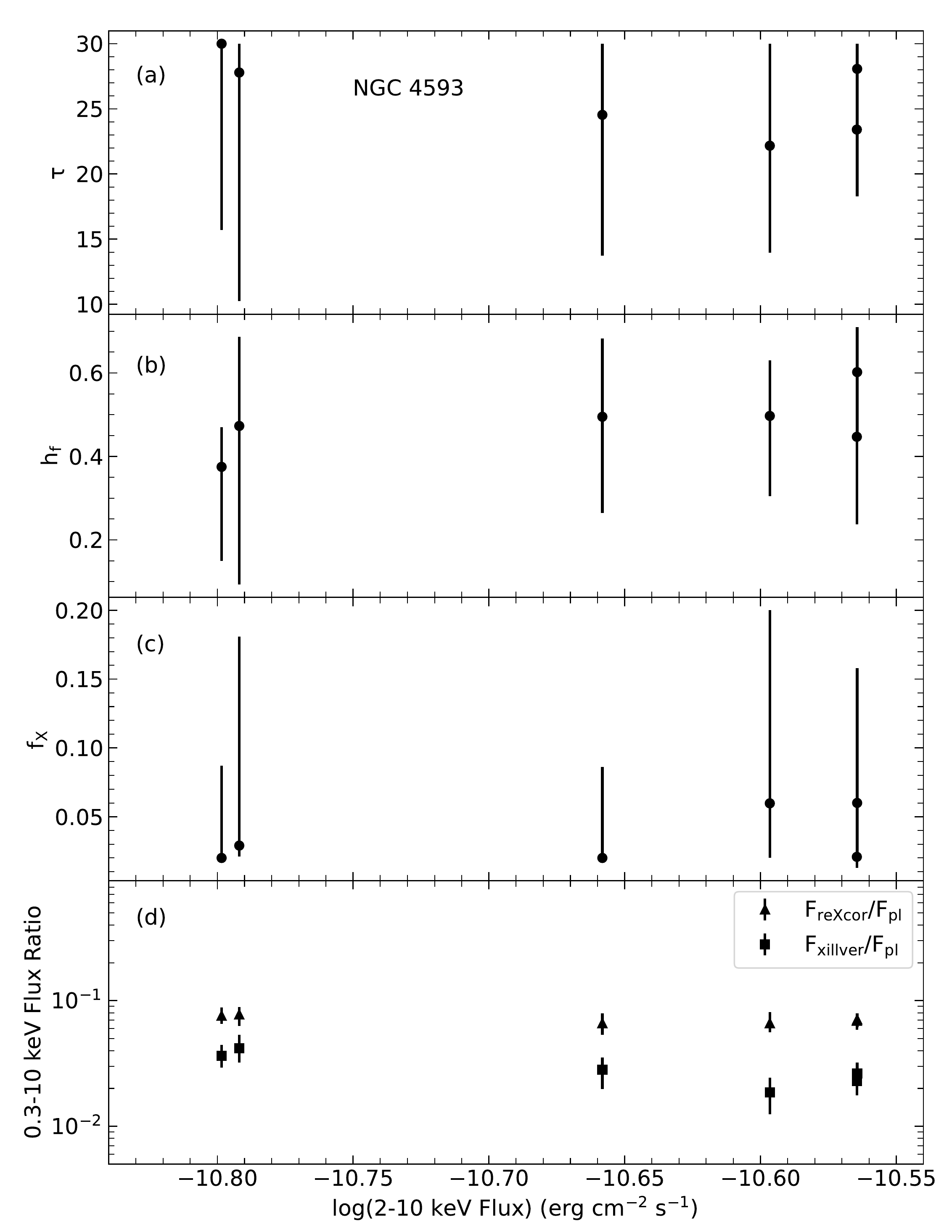}
    \caption{As in Fig.~\ref{fig:heparams}, but now plotting the results from our fits to \ngc\ (Table~\ref{table:ngcfit}). }
    \label{fig:ngcparams}
\end{figure}

The top three panels of Figure~\ref{fig:ngcparams} shows how the three \rexcor\ parameters vary with the observed $2$--$10$~\kev\ flux of \ngc.
Similar to \he, none of the parameters show any correlations with the observed flux. Fig.~\ref{fig:ngcparams}(d) shows the flux ratios of the \rexcor\ and \textsc{xillver} model components relative to the power-law model. The \rexcor\ $0.3$--$10$~\kev\ flux is consistently $\approx 7$\% of the power-law flux, while the \textsc{xillver} flux varies between $2$ and $4$\% of the power-law. This low value of the \rexcor\ flux ratio is largely due to the low ionization state of the inner accretion disc. X-rays absorbed at $\ga 1$~\kev\ will be thermalized and re-emitted at $< 0.3$~keV \citep[e.g.,][]{rfy99}, leading to a reduction in the \rexcor\ flux ratio. As the disc becomes more ionized, fewer hard X-rays can be absorbed and reprocessed in this way. This effect likely explains the lower \rexcor\ ratio found in the \ngc\ fits compared to those for \he\ (Fig.~\ref{fig:heparams}).  

\section{Summary}
\label{sect:summary}
This paper introduces a new phenomenological spectral model of AGNs, \rexcor, that self-consistently combines the effects of a warm corona with the X-ray reflection spectrum from the inner $400$~$r_g$ of an accretion disc. The goal of \rexcor\ is to simultaneously fit both the relativistic reflection signal and the soft excess in AGNs. The model assumes the disc is irradiated by a lamppost X-ray source, and takes into account relativistic light-bending and the ionization gradient on the surface of the disc. To produce a warm corona, accretion energy is injected into the irradiated disc surface, altering the emission and reflection spectrum due to enhanced Comptonization and bremsstrahlung emission. The flux released in the lamppost, the warm corona, and the bulk of the accretion disc must sum to the total local dissipation rate. \rexcor\ spectra can be used to model AGN spectra at energies $\ga 0.3$~\kev.

In this initial release, a total of 8 \rexcor\ table models are available (Table~\ref{table:grids}), separated by specific values of the lamppost height ($h$), the accretion rate ($\lambda$), and the black hole spin ($a$). Each table model contains 20570 \rexcor\ spectra (Table~\ref{table:params}) that are parameterized by the photon-index of the irradiating spectrum ($\Gamma$), the lamppost heating fraction ($f_X$), and the warm corona heating fraction ($h_f$) and Thomson depth ($\tau$). These last three parameters describe changes in the warm corona properties and the distribution of energy in the accretion disc. As a result, varying $h_f$, $f_X$ and $\tau$ lead to wide range of possible soft excess shapes and sizes (Sect.~\ref{sect:rexcor}).

We illustrate the use of \rexcor\ by showing fits to the joint \xmm\ and \nustar\ monitoring campaigns of the Seyfert 1s \he\ and \ngc\ (Sect.~\ref{sect:fits}). The \rexcor\ model provides a good fit to the soft excess in both AGNs with $h_f \approx 0.5$, indicating that a warm corona is an important contributor to the soft excess in both sources. The optical depth of the warm corona is much higher ($\tau \approx 26$) in the low Eddington ratio AGN \ngc\ than in more rapidly accreting \he\ ($\tau \approx 11)$. Examining this relationship, and searching for others, using a wide range of AGNs will lead to new insights into how the energy of the accretion flow is distributed in AGNs. In contrast, it appears to be challenging to use \rexcor\ grids to provide robust constraints on the black hole spin or the height of the lamppost corona without additional information (e.g., spectral-timing analysis from \textit{STROBE-X} observations \citealt{strobex}). However, the derived warm corona parameters of \he\ and \ngc\ are largely insensitive to changes in either $h$ or $a$. Thus, \rexcor\ may be confidently used to determine the warm corona properties of AGNs that lack a black hole spin measurement or an estimate of the coronal height. 

Compelling evidence now exists that the soft excess in AGNs can be explained by the combination of relativistic reflection from the accretion disc with Comptonization in a warm corona \citep[e.g.,][]{xu21}. Systematic use of the \rexcor\ model will allow for a comprehensive test of this idea. \rexcor\ is designed for use with any broadband AGN X-ray spectrum with a good soft X-ray response, including future observations by \textit{XRISM} \citep{xrism}, \textit{Athena} \citep{athena}, and, potentially, \textit{STROBE-X} \citep{strobex}. We expect that the application of \rexcor\ to both archival and future datasets may finally lead to an improved understanding of the soft excess puzzle in AGNs. Future planned releases of \rexcor\ will include a wider range of black hole spins, plus the ability to consider non-Solar abundances.

\section*{Data Availability}
The data underlying this article will be shared on reasonable request to the corresponding author. The \rexcor\ models are publicly available through the XSPEC website.

\section*{Acknowledgements}
X.\ Xiang was supported by a Georgia Tech President's Undergraduate Research Salary Award and a Letson Summer Internship at the School of Physics. The authors thank the International Space Science Institute in Bern, Switzerland for hosting an International Team on "Warm Coronae in AGN". SB acknowledges financial support from ASI under grants ASI-INAF I/037/12/0 and n. 2017-14-H.O. ADR acknowledges financial contribution from the agreement ASI-INAF n.2017-14-H.O.



\bibliographystyle{mnras}
\bibliography{refs} 





\appendix

\section{Additional Examples from the \rexcor\ Grids}
\label{app:a090}
\begin{figure*}
\includegraphics[width=0.8\textwidth]{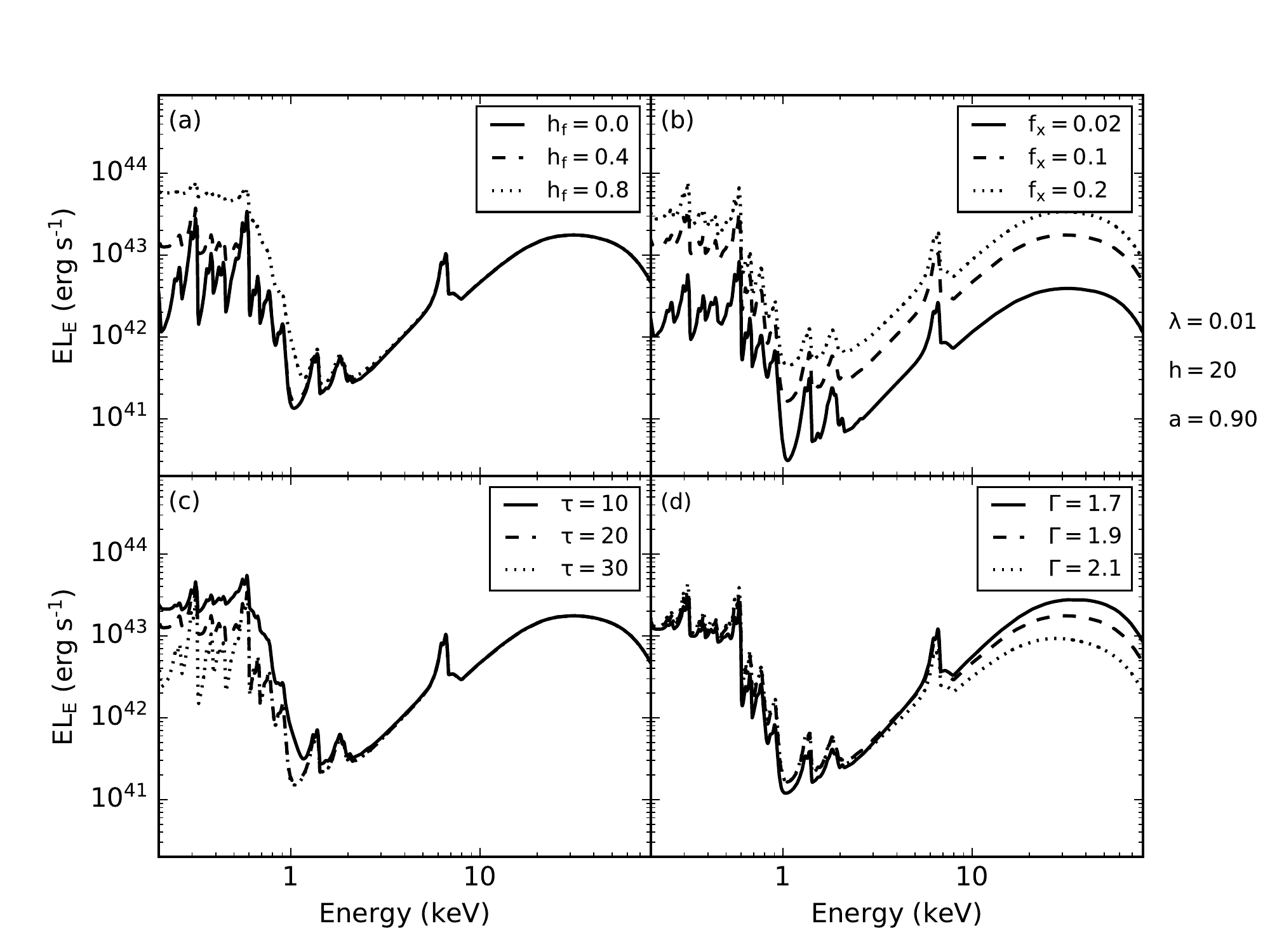}
\includegraphics[width=0.8\textwidth]{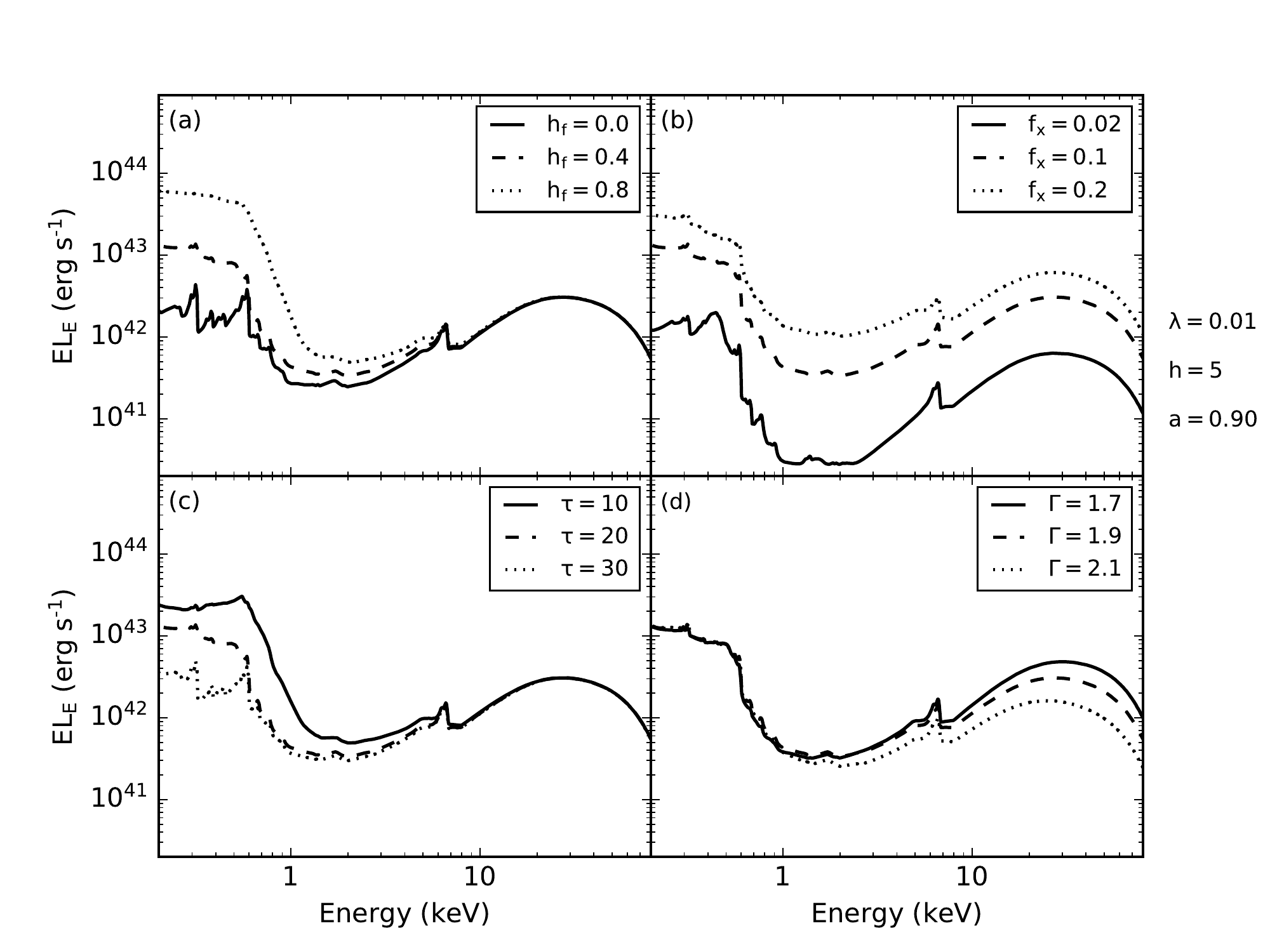}
\caption{As in Fig.~\ref{fig:l01h20a099}, but the spectra are taken from the grid with $\lambda=0.01$ and $a=0.90$. The top figure shows the results with $h=20$ and the lower one plots models with $h=5$. The smaller spin leads to less relativistic blurring when compared to the equivalent $a=0.99$ spectra in Fig.~\ref{fig:l001h20a099}.}
\label{fig:l001h20a090}
\end{figure*}

We present examples of the \rexcor\ model spectra from two of the $a=0.90$ grids. The lower black hole spin increases the radius of the ISCO and reduces the amount of relativistic blurring impacting the model spectra. As a result, the largest impact on the \rexcor\ models is on the spectra that emerge from the inner disc. However, the inner disc is highly ionized when $\lambda=0.1$ (Fig.~\ref{fig:rwarm}), so the largest impact of the lower spin occurs in the $\lambda=0.01$ \rexcor\ models (Fig.~\ref{fig:l001h20a090}.)
Comparing the spectra shown in this figure to the corresponding ones in Fig.~\ref{fig:l001h20a099}, shows that that the lower spin reduces the blurring of the reflection features. In addition, the lower spin somewhat reduces the luminosity of the lamppost (Eq.~\ref{eq:Lx}), which leads to a drop in the ionization state of the disc. Therefore, the $a=0.9$ spectra have a larger contribution from neutral reflection in the final model. As mentioned in Sect.~\ref{sect:rexcor}, the drop in the the \rexcor\ amplitude due to a lower $a$ can be compensated, in part, by increasing $f_X$. This degeneracy limits the ability to use \rexcor\ to constrain black hole spin. The effects of the warm corona parameters (e.g., $h_f$, $\tau$) are unaffected by the lower spin.

\bsp	
\label{lastpage}
\end{document}